\documentclass{article}

\usepackage{amssymb}
\usepackage{amsthm}
\usepackage{amsmath}
\usepackage{subcaption}

\usepackage{xcolor}
\usepackage{lscape}
\usepackage{soul}

\usepackage{graphicx}
\usepackage{hyperref}

\newtheorem{theorem}{Theorem}[section]

\newtheorem{remark}[theorem]{Remark}

\newcommand{\D}{{\mathrm{d}}}

\newcommand{\ic}{{\theta_{11}}}

\begin{document}
\title{Rogue wave formation scenarios for the focusing nonlinear Schr\"odinger equation with parabolic-profile initial data on a compact support}

\author{
F. Demontis$^1$, G. Ortenzi$^{2,4}$, G. Roberti $^3$ and M. Sommacal$^3$ \\
\mbox{} \\
$^1$ Dipartimento di Matematica e Informatica, \\Universit\`{a} degli studi di Cagliari, 09124, Cagliari, Italy\\
$^2$ Dipartimento di Matematica ``Giuseppe Peano'', \\Universit\`{a} di Torino, 10123 Torino, Italy\\
$^3$ Department of Mathematics, Physics and Electrical Engineering, \\Northumbria University, Newcastle upon Tyne, NE1 8ST, UK\\
$^4$ INFN, Sezione di Milano Bicocca
}

\maketitle

\abstract{
We study the $(1+1)$ focusing nonlinear Schr\"{o}dinger equation for an initial condition with compactly-supported parabolic profile and phase depending quadratically on the spatial coordinate.

In the absence of dispersion, using the natural class of self-similar solutions, we provide a criterion for blow-up in finite time, generalising a result by Talanov et al. In the presence of dispersion, we numerically show that the same criterion determines, even beyond the semi-classical regime, whether the solution relaxes or develops a high-order rogue wave, whose onset time is predicted by the corresponding dispersionless catastrophe time. 

The sign of the chirp appears to determine the prevailing scenario among two competing mechanisms for rogue wave formation. For negative values, the numerical simulations are suggestive of the dispersive regularisation of a gradient catastrophe described by Bertola and Tovbis for a different class of smooth, bell-shaped initial data. As the chirp becomes positive, the rogue wave seems to result from the interaction of counter-propagating dispersive dam break flows, as in the box problem recently studied by El, Khamis and Tovbis. 

As the chirp and amplitude of the initial profile are relatively easy to manipulate in optical devices and water tank wave generators, we expect our observation to be relevant for experiments in nonlinear optics and fluid dynamics.
}

\section{Introduction}

In the past decades, much attention has been devoted to the study of the generation mechanisms and of the evolution of rogue waves, that is, unusually high waves, localised in space and time, and substantially higher (typically, more than twice) than the significant wave height of the background on which they emerge. So far, this research has revealed the appearance and relevance of these extreme phenomena in a broad range of different physical contexts, from oceanography \cite{KharifPelinovskySlunyaev2009} to nonlinear optics and lasers \cite{ErkintaloGentyDudley2009}, atmosphere \cite{StenfloShukla2009}, plasma physics \cite{BailungSharmaNakamura2011}, and matter waves (Bose–Einstein condensate) \cite{BludovKonotopoAkhmadiev2009,NguyenDeRandall2017}. In this connection, the focusing nonlinear Schr\"{o}dinger (NLS) equation, featuring quadratic dispersion and cubic nonlinearity of the slowly varying amplitude of just one quasi-monochromatic wave,
\begin{equation}\label{eq:NLS_Foc}
i\epsilon\psi_t+ \frac{\epsilon^2}{2}\psi_{xx}+ |\psi|^2\psi=0\,,
\end{equation}
where $\psi$ is the complex wave envelope and $\epsilon$ is the dispersion parameter defining the modulation scale, plays the role of a universal model \cite{Calogero91,DubrovinGravaKlein2009,BiondiniSitaiMantzavinosTrillo2009,BertolaTovbis2009} as it emerges via multi-scale reduction from almost any conservative wave dynamics \cite{Degasperis2009}, wherever the conditions of weak dispersion, of weak nonlinearity, and of narrow band, are met. Furthermore, as a well-known integrable system \cite{Zakharov1971,ZakharovShabat1972,AblowitzKaupNewellSegur1974,AblowitzPrinariTrubatch2004,CalogeroDegasperis1982}, its Lax pair has been shown to be gauge equivalent to that of other physically relevant models, most famously the Heisenberg ferromagnet \cite{Hasimoto1972,NakamuraSasada1982,DemontisOrtenziSommacal2018}. The NLS equation has been used to provide a simple description of the instability of regular wave trains, a phenomenon nowadays called modulational instability and first predicted by Benjamin and Feir \cite{BenjaminFeir1967} in water dynamics, and by Talanov \cite{Talanov1965} in optics. According to this mechanism, at the linear stage, all perturbations of a plane wave background and with Fourier wave number within a given interval undergo exponential growth: as the instability develops, novel wave patterns, such as rogue waves, have been observed or proved to form \cite{BertolaTovbis2009,AkhmadievPelinovsky2010,OnoratoResidoriBortolozzoMontinaArecchi2013,ElKhamisTovbis2016,GrinevichSantini2018}. The onset and formation of such patterns typically occurs at the nonlinear stage of the evolution, when the linear description associated to the modulational instability mechanism fails.

When one is not concerned with the perturbation of a plane wave of infinite extent, then, as observed in \cite{BonnefoySuretElRandoux2020} and \cite{TikanRobertiElRandouxSuret2022}, two different scenarios for the evolution of unstable wave systems modeled by the NLS equation (\ref{eq:NLS_Foc}) have been considered. In the first scenario, the initial condition is represented by a broad localised wave packet with a smooth envelope, where the nonlinear effects prevail over the exponential growth of small long wave perturbations predicted by the classical modulation instability theory. In this case, Bertola and Tovbis \cite{BertolaTovbis2013} proved that the nonlinear focusing of such smooth wave packets generally leads to a gradient catastrophe that is universally regularised by effect of dispersion, resulting into the emergence of a rogue-wave-like coherent structure, which, in the semiclassical, small-dispersion limit, is asymptotically described by the famous Peregrine soliton solution. In the second scenario, which belongs to the class of the so-called Riemann problems \cite{Biondini2018}, the initial condition features sharp discontinuities in the amplitude, and the resulting evolution is characterised by the formation of dispersive dam break flows \cite{ElKhamisTovbis2016}, the elliptic counterpart of the dispersive shock waves observed in the hyperbolic, defocusing regime \cite{ElHoefer2016}. In particular, in the case of a rectangular barrier \cite{JenkinsMcLaughlin2013}, the so-called `box' or symmetric dam-break problem, in the semiclassical regime, El, Khamis and Tovbis \cite{ElKhamisTovbis2016} showed that the dispersive dam break flows counter-propagating from the edges of the box can interact to form a modulated large-amplitude quasi-periodic breather lattice, whose wave profiles are well approximated by the Akhmediev and Peregrine breathers, thereby suggesting a different mechanism for the formation of rogue waves.

In this paper, we consider a general concave parabolic initial condition on a compact support, which does not immediately fit into one of the two above-mentioned scenarios, but whose evolution shows a prevalence of the underlying rogue-wave generation mechanism from either the first or the second scenario, depending on the value of the chirp and of the amplitude. More specifically, we study (\ref{eq:NLS_Foc}) for the initial condition
\begin{subequations}\label{eq:ini-Talanov}
\begin{equation}\label{eq:ini-Talanov-function}
\psi(x,0)=\sqrt{\mu_0+\gamma_0\,x^2}\,  \exp \hspace{-2pt}\left({\frac{i \alpha_0}{2 \epsilon}x^2} \right)\,,
\qquad\mbox{with}\quad\mu_{0}>0\,,\quad\gamma_{0}<0\,,
\end{equation}
defined on the compact support
\begin{equation}\label{eq:ini-Talanov-support}
x\in\left[-\sqrt{-\frac{\mu_{0}}{\gamma_{0}}},\sqrt{-\frac{\mu_{0}}{\gamma_{0}}}\right]\,.
\end{equation}
\end{subequations}
The endpoints of this compact support are vacuum points, where (\ref{eq:ini-Talanov}) features a catastrophe of the gradient. This problem is of theoretical relevance in the modelling of the interaction of parabolic radio beams with the ionosphere \cite{GurevichShvartsburg1973}. For zero chirp ($\alpha_{0}=0$), the effect of modulational instability to the formation of a region of strong nonlinear oscillations was recently studied in \cite{KamchatnovShaykin2021}. More recently, this initial condition has emerged in the study of mean field games with negative coordination \cite{BonnemainGobronUllmo2020a,BonnemainGobronUllmo2020b}, with applications to crowd dynamics \cite{BonnemainButanoUllmo2023}. The same initial condition -- but for the defocusing NLS equation, which is obtained by replacing the positive sign in front of the nonlinear term in (\ref{eq:NLS_Foc}) with a negative sign -- is well known to appear in the context of the Thomas-Fermi approximation for the Gross-Pitaevskii equation, neglecting quantum pressure and kinetic energy, describing the ground state wave function of a Bose-Einstein condensate confined in a harmonic potential (see \cite{PitaevskiiStringari2016} and references therein).

For the sake of the analysis presented in this paper and in order to introduce our results, we rephrase the NLS equation (\ref{eq:NLS_Foc}) as a fluid dynamics system,
\begin{equation}\label{eq:sys-fluid}
\rho_t+(\rho u)_x=0\,,
\qquad
u_t+uu_x -\rho_x +\epsilon^2 \left( \frac{\rho_x^2}{8 \rho ^2}-\frac{\rho_{xx}}{4 \rho} \right)_x =0\,,
\end{equation}
via the classical Madelung transformation \cite{Madelung1927},
\begin{equation}\label{eq:mapping}
\psi= \sqrt{\rho} \exp \left(\frac{i}{\epsilon} \int {u} \D x \right)\,,
\end{equation}
so that the initial condition (\ref{eq:ini-Talanov}) reads
\begin{equation}\label{eq:ini-Talanov-hydro}
\rho(x,0) = \mu_{0} + \gamma_{0}\,x^{2}\,,
\qquad
u(x,0) = \alpha_{0}\,x\,.
\end{equation}
Using this hydrodynamic form, one can easily identify the dispersive contribution to (\ref{eq:NLS_Foc}) as the terms multiplying $\epsilon^{2}$ in (\ref{eq:sys-fluid}). Therefore, a dispersionless counterpart of (\ref{eq:NLS_Foc}) can be constructed by setting $\epsilon$ to zero in (\ref{eq:sys-fluid}), obtaining the quasi-linear system of geometric optics equations:
\begin{equation}\label{eq:ellsys-fluid}
\rho_t+u \rho_x + \rho u_x=0\,,
\qquad
u_t+uu_x-{\rho_x}  =0\,.
\end{equation}
This system is called elliptic as its associated ``characteristic velocities'' are complex.

As the dispersionless counterpart of the NLS equation, system (\ref{eq:ellsys-fluid}) has been the object of intense research since its introduction in the context of nonlinear optics and plasma physics (nonlinear interaction of radio-waves with ionosphere), starting from the seminal contributions by Talanov \cite{Talanov1965,Talanov1970}, Gurevich and Shvartsburg \cite{GurevichShvartsburg1970,GurevichShvartsburg1973}, and Akhmanov, Sukhorukov and Khokhlov \cite{AkhmanovSukhorukovKhokhlov1970,AkhmanovSukhorukovKhokhlov1968} (see also \cite{KamchatnovShaykin2021} and references therein).

The elliptic system (\ref{eq:ellsys-fluid}) has also been consistently investigated as the semi-classical limit of the NLS equation in \cite{BronskiKutz1999,CaiMcLaughlinMclaughlin2002,CenicerosTian2002,ClarkeMiller2002,LeeLyngVanakova2012}.

In particular, the approach followed in \cite{GurevichShvartsburg1970} is based on the introduction of a so-called hodographic transform (see \cite{LandauLifshitzVI}), which converts the nonlinear system (\ref{eq:ellsys-fluid}) in a linear system for the inverse functions $x(\rho,u)$ and $t(\rho,u)$,
\begin{equation}\label{eq:ellsys-fluid-hodo}
x_u-u t_u+\rho t_\rho=0\,,
\qquad
t_u+ut_\rho-x_\rho=0\,.
\end{equation}

However, for initial conditions of the form (\ref{eq:ini-Talanov-hydro}), Talanov in 1965 \cite{Talanov1965,Talanov1970,GurevichShvartsburg1970,GurevichShvartsburg1973} observed that system \eqref{eq:ellsys-fluid} admits solutions where $\rho$ and $u$ are polynomials in the $x$ variable. For zero chirp ($\alpha_{0}=0$) and for $\rho(x,0)=1-x^{2}$, the solution was shown to feature a blow-up at finite time $t_{c}=\frac{\pi}{4}$. For positive phase ($\alpha_{0}>0$) and for $\rho(x,t)=\mu(t)+\gamma(t)\,x^{2}$ with $\gamma(t)=-\mu^{3}(t)$, a condition on $\alpha_{0}$ was derived (see \cite{GurevichShvartsburg1973}), dictating whether the dispersionless solution features asymptotic relaxation ($\alpha_{0}>2$) or blow-up at finite time ($\alpha_{0}<2$), and an expression for the time of catastrophe was provided. The evolutions studied by Talanov, Gurevich and Shvartsburg belong to the natural class of the so-called self-similar solutions which are particularly useful to describe metastable dynamics \cite{BarenblattZeldovich1972}. Analogous self-similar polynomial solutions were introduced by Ovsyannikov in 1979 \cite{Ovsyannikov1979} and studied in \cite{BrazhnyiKamchatnovKonotop2003,CamFalOrtPed2019,CamFalOrtPedTho2019} for the hyperbolic case, corresponding to the defocusing NLS equation, showing that an asymptotic relaxation behavior is always present for any initial chirp, differently from the elliptic case. The possible relevance -- to which the present work brings potential evidence in support -- of the self-similar solutions introduced by Talanov to the study of the full focusing NLS equation in the case of small dispersion has been recently pointed out in \cite{Suleimanov2017}.

Continuing on the line of Talanov's approach, in this work we provide closed-form, implicit expressions for the dispersionless solutions of (\ref{eq:ellsys-fluid}) with (\ref{eq:ini-Talanov-hydro}) for a generic concave parabolic profile (to ensure the compactness of the support) and for generic chirp $\alpha_{0}$ (note that the case $\alpha_{0}<0$ cannot be obtained from the case $\alpha_{0}>0$ by a rescaling): analogously to \cite{GurevichShvartsburg1973}, the dispersionless solutions either undergo asymptotic relaxation or blow-up at finite time $t_{c}$, depending on the sign of the quantity $\alpha_{0}^{2}+4\,\gamma_{0}$. The time of blow-up can be explicitly given in terms of the parameters $\alpha_{0}$ and $\gamma_{0}$ and is independent from $\mu_{0}$. 

More significantly, we numerically show the notable fact that, in a dispersion regime that extends even beyond the semi-classical approximation, the solutions to the fully dispersive NLS equation (\ref{eq:NLS_Foc}) with (\ref{eq:ini-Talanov}) seem to be driven by the corresponding solutions of the dispersionless counterpart (\ref{eq:ellsys-fluid}) with (\ref{eq:ini-Talanov-hydro}): the dispersive solutions feature asymptotic relaxation when the dispersionless counterpart relaxes, whereas they develop a space-time-localised peak in the form of a Peregrine-like rogue-wave when the dispersionless counterpart features a catastrophe. In fact, the amplitude of the central peak, as well as the number and distribution of the side peaks, suggest that we are observing a high-order rogue-wave solution (see \cite{BilmanLingMiller2020, BilmanMiller2022, KedzioraAnkiewiczAkhmediev2011, HeZhangWangPorsezianFokas2013,LingZhang2022-1,LingZhang2022-2} and references therein). Similarly to what is analytically found in (\cite{BertolaTovbis2013}), when the conditions allow for the formation of a localised peak, this latter occurs at a time that is very well approximated by the time of the catastrophe of the corresponding dispersionless solution. In particular, although our initial condition (\ref{eq:ini-Talanov}) does not fit the class of analytic, bell-shaped initial data studied in \cite{BertolaTovbis2013}, for $\alpha_{0}^{2}+4\,\gamma_{0}<0$ the emergence of a (high-order) rogue wave in the numerical evolution of the initial condition shows a remarkable similarity with what the one described in \cite{BertolaTovbis2009,BertolaTovbis2013,GrimshawTovbis2013}, where Peregrine breathers emerge during the regularisation of the generic gradient catastrophe (Bertola-Tovbis scenario). However, as $\alpha_{0}^{2}+4\,\gamma_{0}$ increases and becomes larger than zero, the evolution is more and more characterised by the formation of a lattice of counter-propagating breather-like structures with a central higher peak produced by constructive nonlinear interference, resembling what is observed in the context of the dam break problem studied in \cite{ElKhamisTovbis2016} (El-Khamis-Tovbis scenario)), again in spite of the substantial analytical differences among our initial condition and theirs.

Although we currently do not have an analytical explanation of our numerical observations, following the above analogies, we conjecture that the mechanism, in the evolution of (\ref{eq:ini-Talanov}) in the small-dispersion semiclassical regime, leading to the generation of a space-time-localised peak, when the solution does not undergo asymptotic relaxation, may be given by the interplay of the Bertola-Tovbis and El-Khamis-Tovbis scenarios, with the Bertola-Tovbis scenario prevailing when $\alpha_{0}^{2}+4\,\gamma_{0}<0$ and with the El-Khamis-Tovbis scenario prevailing when $\alpha_{0}^{2}+4\,\gamma_{0}>0$.

This paper is structured in two parts. In Section \ref{sec:ansol}, we provide a complete classification and characterisation of compactly supported Talanov solutions of the dispersionless model and we identify the criterion for their blow-up. The results are summarised in Appendix \ref{sec:appendix}. In Section \ref{sec:num}, we compare the semi-analytical solution of the dispersionless model to the full focusing NLS equation, testing the criterion for the focusing blow-up against the formation of a rogue-wave-like peak for the focusing NLS equation. The numerics shows that the dispersionless blow-up is related to the appearance of a rogue wave in the full dispersive equation.

\section{Study of the Talanov solutions}
\label{sec:ansol}
We approach the study of compactly supported self-similar solutions of the dispersionless focusing NLS equation by generalising what has been done in \cite{Talanov1965,GurevichShvartsburg1970} (for the hyperbolic analogue, see \cite{Ovsyannikov1979,CamFalOrtPed2019}).

We start by considering the Ansatz
\begin{equation}\label{eq:polsol}
\rho=\gamma(t) x^2+ \omega(t) x + \zeta(t)\,,\qquad
u=\alpha(t) x+\beta(t)\, .
\end{equation}

\begin{remark}
This Ansatz, due to Talanov, has also the remarkable property that it is the only truncation of the expansions $\rho=\sum_{k=0}^\infty \rho_{k}(t)x^k$ and $u=\sum_{k=0}^\infty u_{k}(t)x^k$ leading to an exact solution.\end{remark}

From (\ref{eq:ellsys-fluid}), the coefficients satisfy
\begin{equation}
\label{eq:5ODE}
\gamma'+3\alpha \gamma=0\,,\qquad
\alpha'+\alpha^2-2\gamma=0\, ,\qquad
\omega'+2 \alpha \omega +2 \beta \gamma=0\,,\qquad
\zeta'+\alpha \zeta +\beta \omega=0\,,\qquad
\beta' + \alpha \beta-\omega=0\, ,
\end{equation}
where a prime appended to a symbol indicates differentiation with respect to $t$, $f'\equiv \D f/\D t$.

A convenient change of variables for the above system can be had by looking at the geometrical properties of $\rho$ in (\ref{eq:polsol}). Indeed, considering $\rho$ as a parabola in the spatial coordinate $x$, let $\mu$ and $\xi$ be its vertex height and position, respectively. We have
\begin{equation}\label{eq:parabola_height_position}
\mu = \zeta-\omega^2/(4 \gamma)\,,\qquad
\xi=  -\omega/(2 \gamma)\,,
\end{equation}
which, using (\ref{eq:5ODE}), can be shown to satisfy the following ordinary differential equations:
\begin{equation}\label{eq:parabola_height_position_diff}
\mu'+\alpha \mu=0\,,\qquad
\xi'=\beta-\alpha \xi\,,\qquad
\xi''=0\,.
\end{equation}
These relations imply that the quantity $\beta-\alpha \xi$ is constant. Consequently, system (\ref{eq:5ODE}) can be rewritten as a system of three differential equations in the three unknowns $\alpha$, $\gamma$ and $\mu$,

\begin{equation}
\label{eq:3ODE-sym}
\gamma'+3\alpha \gamma=0\,,\qquad
\alpha'+\alpha^2-2\gamma=0\,,\qquad
\mu'+\alpha \mu=0\,.
\end{equation}
We observe that this is the system related to the symmetric solution obtained by setting $\omega$ and $\beta$ to zero and $\zeta=\mu$ in (\ref{eq:5ODE}), corresponding to the Ansatz
\begin{equation}\label{eq:symsol}
\rho=\gamma(t) x^2+ \mu(t)\,,\qquad
u=\alpha(t) x\,.
\end{equation}
The dispersive counterpart of this evolution is generated by the initial datum (\ref{eq:ini-Talanov}),
\begin{equation}
\psi(x,0)=\sqrt{\mu_0+\gamma_0 x^2}\,  \exp \hspace{-2pt}\left({\frac{i \alpha_0}{2 \epsilon}x^2} \right)\,,
\end{equation}
where
\begin{equation}
\label{eq:init-3ODE-sym}
\gamma(0)=\gamma_{0}\,,\qquad
\alpha(0)=\alpha_{0}\,,\qquad
\mu(0)=\mu_{0}\,.
\end{equation}

Note that system (\ref{eq:3ODE-sym}) coincides with the system obtained in the hyperbolic case studied in \cite{CamFalOrtPed2019}, with the except of an opposite sign in front of the term $2\gamma$ in the second ODE in the hyperbolic case. As we will see in the following, this sign difference is responsible for many new interesting phenomena in the elliptic case.

The exponential nature of $\mu$ and $\gamma$ can be easily seen from their evolution equations \eqref{eq:3ODE-sym}. Indeed, by integrating the first and the third equation in system \eqref{eq:3ODE-sym}, we get
\begin{equation}\label{eq:expevol}
\gamma(t) = \gamma_0 \exp \hspace{-2pt}\left(- 3 \int_0^t \alpha(\tilde{t})\, \D \tilde{t}   \right)  \, ,\qquad
\mu(t) = \mu_0 \exp\hspace{-2pt}\left(-\int_0^t \alpha(\tilde{t})\, \D \tilde{t}\right)\, .
\end{equation}
Therefore, $\gamma$ and $\mu$ cannot change sign during their evolution. The fact that $\mu$ stays positive if it is positive at $t=0$ is physically relevant as it ensures the positivity of $\rho= |\psi|^2$. The positivity of $\rho$ is also mathematically relevant because, if $\rho<0$, then system (\ref{eq:ellsys-fluid}) changes its nature, becoming hyperbolic.

Let us now solve system (\ref{eq:3ODE-sym}). First of all, after a technical change of variables (see \cite{CamFalOrtPed2019}),
\begin{equation}\label{eq:par-def}
\sigma=(\gamma/\gamma_0)^{1/3}\,,
\qquad \mbox{entailing}\qquad
\sigma(0)=1\,,
\end{equation}
we restrict our attention to the closed subsystem
\begin{equation}\label{eq:minimal}
\alpha'+\alpha^2-2\gamma_0 \sigma^3=0\,,\qquad
\sigma'+\alpha \sigma=0\, .
\end{equation}
After multiplying the first equation in (\ref{eq:minimal}) by $\alpha$, rewriting it as an equation for the variable $\alpha^{2}$, and finally combining it with the second equation for $\sigma'$, we obtain a single linear equation in $\alpha^2$ as a function of $\sigma$,
\begin{equation}\label{eq:alpha2_sigma}
\frac{\D \alpha^2}{\D \sigma} -  \frac{2}{\sigma}\alpha^2+4 \gamma_0 \sigma^2=0\,,
\end{equation}
whose solution is
\begin{subequations}\label{eq:alpha_sigma}
\begin{equation}\label{eq:alpha_sigma_eq}
\alpha(\sigma) = \pm \sqrt{(\alpha_0^2+4\gamma_0) \sigma^2-4 \gamma_0 \sigma^3}\,  .
\end{equation}
The choice of the sign is determined by the sign of $\alpha(\sigma(0))$, \begin{equation}\label{eq:alpha_sigma_init}
\alpha\big(\sigma(0)\big)=\left.\alpha\right|_{t=0}=\left.\alpha\right|_{\sigma=1}=\alpha_0\,.
\end{equation}
\end{subequations}
For the sake of completeness, we add here the dependence on $\sigma$ of the parabola curvature $\gamma$ and the parabola vertex $\mu$,
\begin{equation} \label{eq:curvmaxpar}
\gamma=\gamma_0 \sigma^3\, ,\qquad
\mu=\mu_0 \sigma \,,
\end{equation}
which have been obtained by combining (\ref{eq:expevol}) with the definition of $\sigma$ in (\ref{eq:par-def}).

When $\alpha_0=0$, system (\ref{eq:minimal}) suggests that the sign
{ of (\ref{eq:alpha_sigma_eq})} is determined by the sign of $\gamma_0$
{ and consequently the ODE for $\sigma(t)$ is}
\begin{equation}\label{eq:curvODE}
\sigma' \pm\sqrt{(\alpha_0^2+4\gamma_0) \sigma^4-4 \gamma_0 \sigma^5}=0\, ,
\qquad\mbox{with}\quad \sigma(0)=1\, .
\end{equation}

In the rest of this section and of the following one, we focus on Gurevich-Shvartsburg-type solutions \cite{GurevichShvartsburg1970,GurevichShvartsburg1973}, namely solutions to (\ref{eq:ellsys-fluid}) for an initial condition (\ref{eq:ini-Talanov-hydro}) where $\gamma_0<0$ and the phase coefficient $\alpha_0$ is generic. As shown in \cite{GurevichShvartsburg1970}, in the context of nonlinear optics, the relevance of these solutions resides in their self-similarity. Indeed, they can be seen as the largest self-similar generalisation of
\begin{equation}
\rho= -\left( \frac{x}{3t} - c\right)^2  \, , \qquad
u= 2 \frac{x}{3t} + c\, ,
\label{eq:rarell}
\end{equation}
which is the elliptic counterpart of the rarefaction solution for the dam break problem or the gas dynamics problem with pressure proportional to $\rho^2$ (\textit{e.g.}, see \cite{Whitham}).
We remark that the solution (\ref{eq:rarell}) is fundamental in the defocusing/hyperbolic case, whereas it is unphysical in the focusing/elliptic case because $\rho$ is negative. In fact, in the focusing/elliptic case there is no physically relevant solution with self-similarity variable $\xi=x/t$. Therefore, the Ansatz (\ref{eq:symsol}) is the only way to analyse generic structures of selfsimilar solutions for (\ref{eq:ellsys-fluid}).

Such solutions live on the compact support $\mathcal{D}$ given by
\begin{equation}\label{eq:support}
\mathcal{D}=\{ x \in \mathbb{R} :  |x|<\sqrt{-\frac{\mu_0}{\gamma_0}\, }\,  \frac{1}{\sigma}\}\, .
\end{equation}

Before moving onto the classification of the dispersionless dynamics, it is worth noting that system (\ref{eq:3ODE-sym}) admits also solutions with $\gamma_0>0$, \textit{i.e.} parabolas with positive concavity corresponding to a wave function with a noncompact support. Contrary to the defocusing/hyperbolic case, the nonsmooth gluing of solutions, necessary to restore a physically relevant solution, cannot be achieved by standard characteristics methods \cite{Whitham}. It seems that  solutions with positive concavity could display quite different phenomena requiring  a separate  study.

In order to simplify the classification of the solutions, we introduce the parameters:
\begin{equation}\label{eq:paramAB}
A = \alpha_0^2+4\gamma_0\, , \qquad B = -4\gamma_0>0\, .
\end{equation}
In the following, using $A$ and $B$, we complete the characterisation of the compactly-supported parabolic solutions: our results are summarised in Appendix \ref{sec:appendix}.

\subsection{Nonnegative chirp $\alpha_0 \geq 0$}
\begin{figure}[!h]
\centering
\begin{subfigure}{0.23\textwidth}
  \centering
  \includegraphics[width=\linewidth]{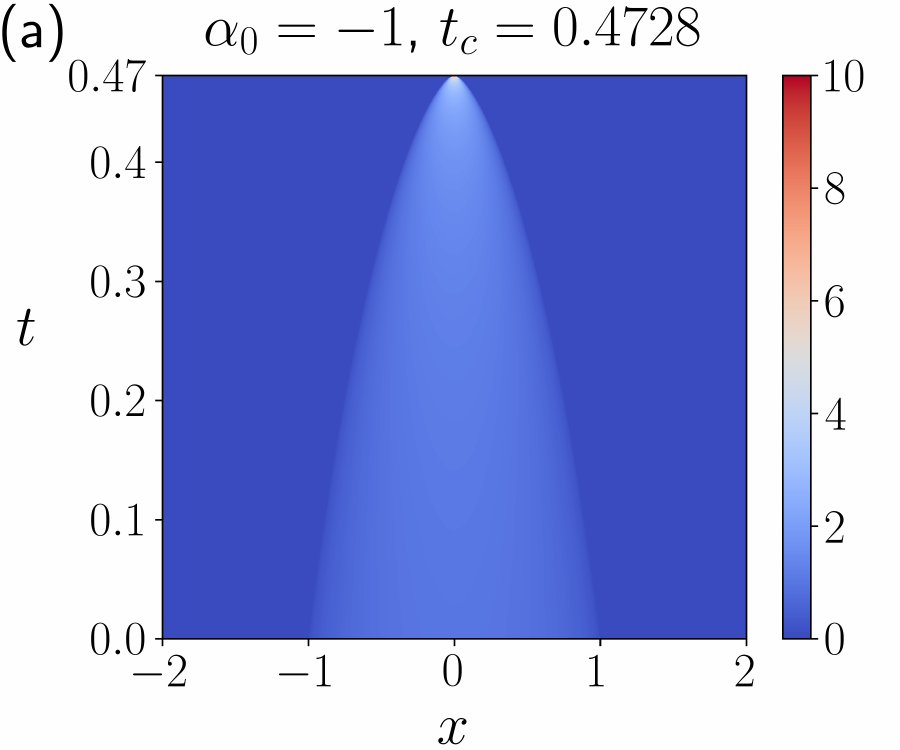}
\end{subfigure}\,\,\,\,
\begin{subfigure}{0.23\textwidth}
  \centering
  \includegraphics[width=\linewidth]{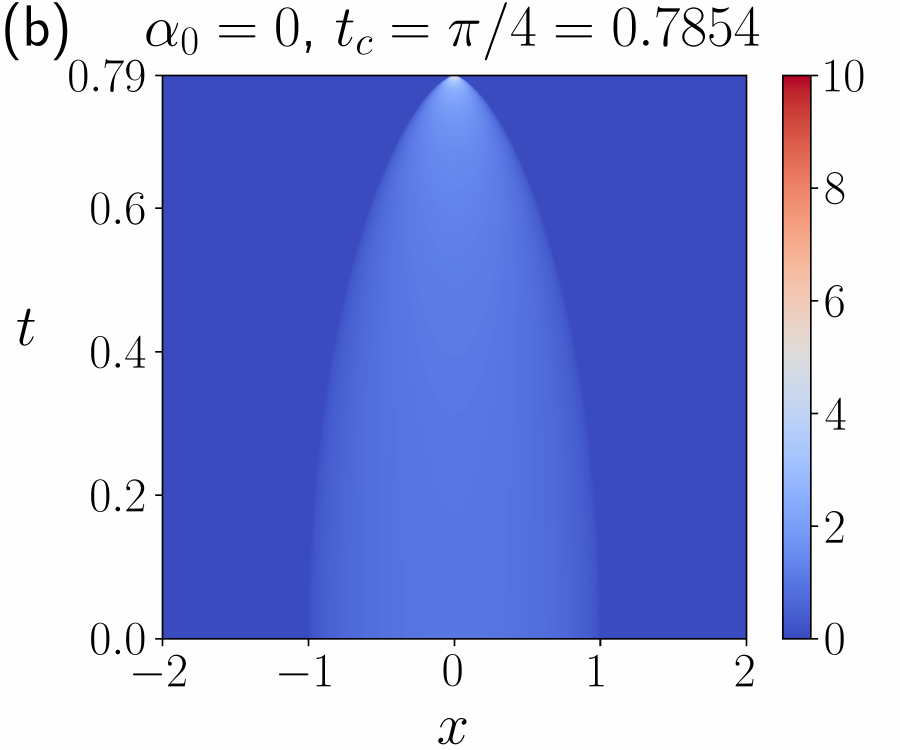}
\end{subfigure}\,\,\,\,
\begin{subfigure}{0.23\textwidth}
  \centering
  \includegraphics[width=\linewidth]{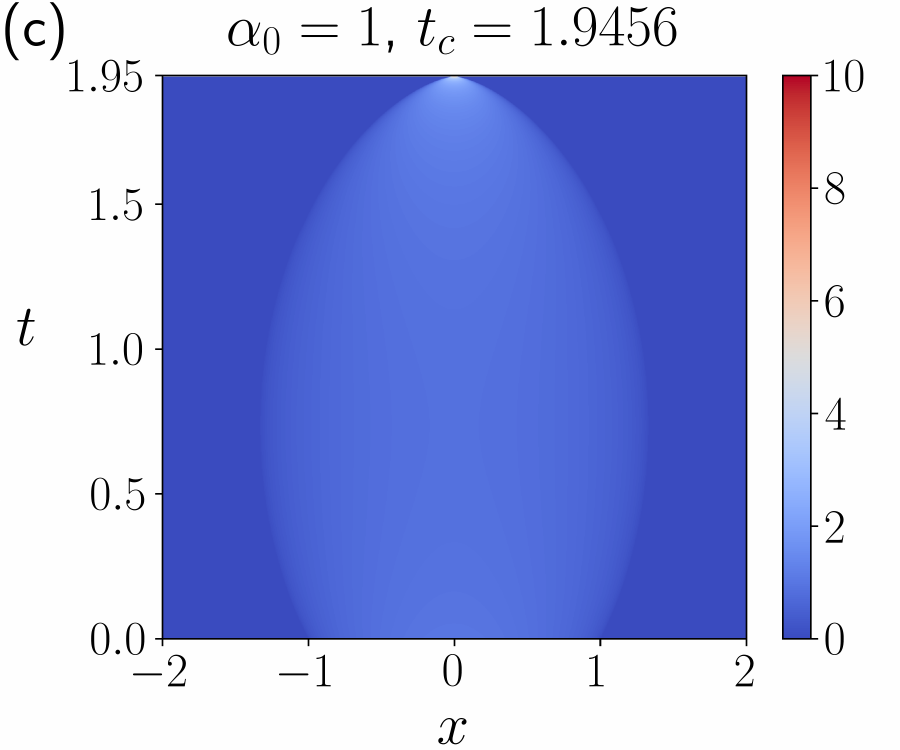}
\end{subfigure}\,\,\,\,
\begin{subfigure}{0.23\textwidth}
  \centering
  \includegraphics[width=\linewidth]{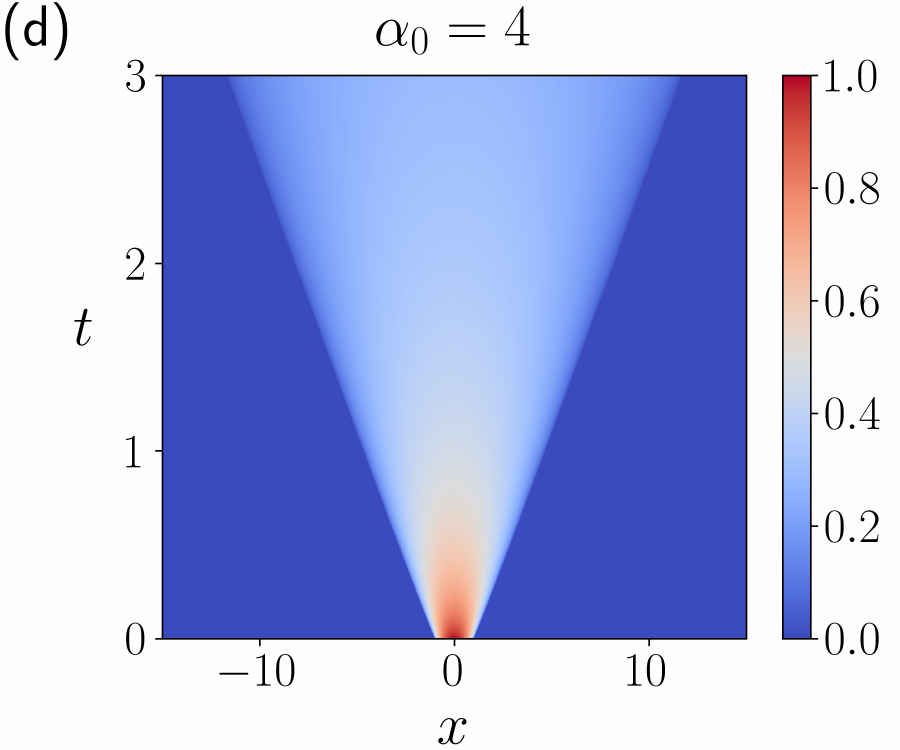}
\end{subfigure}
\caption{Spatio-temporal evolution of $\sqrt{\rho}$, the absolute value of the solution to (\ref{eq:ellsys-fluid}), emerging from the initial condition (\ref{eq:ini-Talanov}) with $\gamma_{0}=-1$ and $\mu_{0}=1$, for different values of $\alpha_{0}$. With this choice of $\gamma_{0}$ and $\mu_{0}$, the dispersionless solution is expected to feature a blow-up in finite time $t_{c}$ for $\alpha_{0}<2$ and to undergo asymptotic relaxation for $\alpha_{0}\geq2$.}
\label{Fig:plot_dispersionless}
\end{figure}

When $\alpha_0>0$, for small times $\sigma$ decreases, i.e. $\sigma'(0^+)<0$. We have the following three distinct cases, depending on the sign of $A$.
\begin{enumerate}
\item\label{case:A>0} %
    When $A>0$, namely, when $\alpha_0^2+4\gamma_0>0$, recalling that the parameter $\sigma$ defined in (\ref{eq:par-def}) is always positive, the solution of (\ref{eq:3ODE-sym}) reads (\cite{GradshteynRyzhik}, 2.228.1 p.86)
    \begin{equation}\label{eq:sol-par}
    \begin{split}
    t =&
    \frac{\sqrt{A+B \sigma}}{A\sigma }
    -\frac{\sqrt{A+B }}{A }
    +
    \frac{B}{2 A^{3/2}} \log \left[ \frac{(\sqrt{A+B}+\sqrt{A})(\sqrt{A+B \sigma}-\sqrt{A})}{(\sqrt{A+B}-\sqrt{A})(\sqrt{A+B \sigma}+\sqrt{A})}\right]
    \, .
    \end{split}
    \end{equation}
    The auxiliary variable $\sigma$ is monotonic in time and approaches $0$ for large $t$. Thus, the amplitude $\sqrt{\rho}$ features asymptotic relaxation over time.
\item\label{case:A=0}%
    When $A=0$, namely, when $\alpha_0^2+4\gamma_0=0$, the solution (\ref{eq:sol-par}) degenerates and becomes
    \begin{subequations}\label{eq:sol-par-0}
    \begin{equation}
    t=\frac{1-\sigma ^{3/2}}{3 \sqrt{-\gamma _0} \sigma ^{3/2}}
    \end{equation}
    which can be explicitly inverted  giving
    \begin{equation}
    \sigma=\frac{1}{\left(1+3 \sqrt{-\gamma_0}\, t \right)^{2/3}}\, .
    \end{equation}
    \end{subequations}
    Also in this case, $\sigma$ approaches zero asymptotically in time, and the amplitude $\rho$ undergoes relaxation.
    {In this case is possible to write explicitly the solution
    \begin{equation}
    \rho=\frac{\mu_0}{\left(1+3 \sqrt{-\gamma_0}\, t \right)^{2/3}}-|\gamma_0| \left(\frac{x}{1+3 \sqrt{-\gamma_0}\, t} \right)^2\, , \qquad
    u=  2 \sqrt{-\gamma_0} \frac{ x}{1+3 \sqrt{-\gamma_0} t}\, .
    \label{eq:simplerelax}
    \end{equation}
    Even if  (\ref{eq:simplerelax}) seems similar to (\ref{eq:rarell}),  now the term of zeroth order in $x$  is time dependent. This crucial difference changes the time exponent of the self-similarity variable, giving  $\xi \sim x/\left(1+3 \sqrt{-\gamma_0}\, t \right)^{2/3}$.}
\item\label{case:A<0}
    When $A<0$, namely, when $\alpha_0^2+4\gamma_0<0$, differently from the previous two cases, the auxiliary variable $\sigma$ is not monotonic in time. For small times, $\sigma < 1$ and the implicit solution takes the form
    \begin{equation}\label{eq:sol-par-1cat}
    \begin{split}
    t =&
    \frac{\sqrt{A+B \sigma}}{A \sigma }-
    \frac{B\, }{(-A)^{3/2}}  \mathrm{arctan} \left(\frac{\sqrt{A+B \sigma}}{\sqrt{-A}}\right)
    -\frac{\sqrt{A+B }}{A }+
    \frac{B\, }{(-A)^{3/2}}  \mathrm{arctan} \left(\frac{\sqrt{A+B }}{\sqrt{-A}}\right)
    \,,
    \end{split}
    \end{equation}
    where $A+B>0$ by definition (\ref{eq:paramAB}). The argument of the square roots in (\ref{eq:sol-par-1cat}) vanishes at the minimum $\sigma_{\mathrm{min}}$ of $\sigma(t)$ (\textit{i.e.}, where $\sigma'=0$),
    \begin{subequations}\label{eq:minimum}
    \begin{equation}\label{eq:s-min}
    \sigma_{\mathrm{min}} \equiv -\frac{A}{B}=1-\frac{\alpha_0^2}{4|\gamma_0|} >0
    \end{equation}
    which corresponds to the time
    \begin{equation}
    \begin{split}\label{eq:t-inv}
    t_{\mathrm{min}} =&
     -\frac{\sqrt{A+B }}{A }+
    \frac{B\, }{(-A)^{3/2}}  \mathrm{arctan} \left(\frac{\sqrt{A+B }}{\sqrt{-A}}\right)
    \,  .
    \end{split}
    \end{equation}
    \end{subequations}
    As $\sigma''(t_{\mathrm{min}})>0$, for $t>t_{\mathrm{min}}$ the auxiliary variable $\sigma$ starts to increase monotonically, $\sigma'(t)>0$ for all $t>t_{\mathrm{min}}$. It is straightforward to see that there are no other critical points, and hence we have to solve equation (\ref{eq:curvODE}) with the minus sign
    {in front of the square root}, obtaining
    \begin{equation}\label{eq:sol-par-2cat}
    \begin{split}
    t =&
    - \frac{\sqrt{A+B \sigma}}{A \sigma }+
    \frac{B\, }{(-A)^{3/2}}  \mathrm{arctan} \left(\frac{\sqrt{A+B \sigma}}{\sqrt{-A}}\right)
    -\frac{\sqrt{A+B }}{A }+
    \frac{B\, }{(-A)^{3/2}}  \mathrm{arctan} \left(\frac{\sqrt{A+B }}{\sqrt{-A}}\right)
    \, ,
    \end{split}
    \end{equation}
    which is defined on the semi-open interval $[0,t_{c})$, with
    
    \begin{equation}\label{eq:catatime}
    t_c=-\frac{\sqrt{A+B }}{A }+
    \frac{B\, }{(-A)^{3/2}}  \left[\frac{\pi}{2}+\mathrm{arctan} \left(\frac{\sqrt{A+B }}{\sqrt{-A}}\right) \right]
    \, ,
    \end{equation}
    obtained by taking the limit $\sigma \to \infty$ in (\ref{eq:sol-par-2cat}). The solution (\ref{eq:sol-par-2cat}) is not defined for $t\geq t_{c}$. Correspondingly the amplitude $\rho$ features a blow-up at the finite time $t=t_{c}$.
\end{enumerate}
In Figure \ref{Fig:plot_dispersionless}, we show three examples of blow-up at finite time (for $A<0$) and one example of relaxation behaviour (for $A>0$) for the dispersionless solution emerging from the initial condition (\ref{eq:ini-Talanov}) with $\gamma_{0}=-1$ and $\mu_{0}=1$.

\begin{remark}
The case $\alpha_0=0$ (particular subcase of the $A<0$ case) has already been studied in \cite{GurevichShvartsburg1970}: the solution blows-up at $t_c=\frac{\pi}{4 \sqrt{|\gamma_0|}}$ and $t_{\mathrm{min}}=0$.
\end{remark}

\subsection{Negative chirp $\alpha_0 < 0$}
When  $\alpha_0<0$, we have $\sigma'>0$ and the sign of $A$ cannot prevent the blow-up. The behavior for different values of $A$ is qualitatively similar, and the motion of the maximum of the density is monotonic in time. The computations are the same as in the previous section and we limit ourselves to merely listing the results.

The solution in this case is, for $A<0$,
\begin{equation}
\begin{split}
t =&
- \frac{\sqrt{A+B \sigma}}{A \sigma }+
\frac{B\, }{(-A)^{3/2}}  \mathrm{arctan} \left(\frac{\sqrt{A+B \sigma}}{\sqrt{-A}}\right)
+\frac{\sqrt{A+B }}{A }-
\frac{B\, }{(-A)^{3/2}}  \mathrm{arctan} \left(\frac{\sqrt{A+B }}{\sqrt{-A}}\right)
\,  ,
\end{split}
\label{eq:sol-par-phaseneg-1}
\end{equation}
until the blow-up time
\begin{equation}
 t_c=\frac{\sqrt{A+B }}{A }+
\frac{B\, }{(-A)^{3/2}}  \left[\frac{\pi}{2} -\mathrm{arctan} \left(\frac{\sqrt{A+B }}{\sqrt{-A}}\right) \right]
\,  .
\label{eq:catatime-phaseneg-1}
\end{equation}
In the degenerate case $A=0$, \textit{i.e.} $\alpha_0=-2\sqrt{-\gamma_0}$ for $\alpha_0<0$, we have the solution
\begin{equation}
\sigma=\frac{1}{\left(1-3 \sqrt{-\gamma_0}\, t \right)^{2/3}}\, ,
\end{equation}
until the blow-up time
\begin{equation}
t_c=\frac{1}{3 \sqrt{-\gamma_0}}\, .
\end{equation}
We complete the classification with the case $A>0$, where the solution is
\begin{equation}
\begin{split}
t =&
-\frac{\sqrt{A+B \sigma}}{A\sigma }
+\frac{\sqrt{A+B }}{A }
-
\frac{B}{2 A^{3/2}} \log \left[ \frac{(\sqrt{A+B}+\sqrt{A})(\sqrt{A+B \sigma}-\sqrt{A})}{(\sqrt{A+B}-\sqrt{A})(\sqrt{A+B \sigma}+\sqrt{A})}\right]
\, ,
\end{split}
\label{eq:sol-bu-phaseneg2}
\end{equation}
until the blow-up time
\begin{equation}
\begin{split}
t _c=&
\frac{\sqrt{A+B }}{A }
-
\frac{B}{2 A^{3/2}} \log \left( \frac{\sqrt{A+B}+\sqrt{A}}{\sqrt{A+B}-\sqrt{A}}\right)
\, .
\end{split}
\label{eq:t-bu-phaseneg2}
\end{equation}

In Figure \ref{Fig:plot_tc_vs_alpha0} we show the dependence of the blow-up time on the phase parameter $\alpha_0$, for the initial datum $\gamma_0=-1$, $\mu_0=1$ in the initial data class (\ref{eq:ini-Talanov}).
\section{Dispersive effects on Talanov initial data for the focusing NLS equation}
\label{sec:num}

The goal of the present section is to qualitatively compare the numerical solution of the full dispersive NLS equation (\ref{eq:NLS_Foc}) with the analytic solution of its dispersionless counterpart (\ref{eq:ellsys-fluid}) given in the previous section. Here we consider the initial condition (\ref{eq:ini-Talanov}) for both the dispersive and the dispersionless model, with $\mu_{0}=1$ and $\gamma_{0}=-1$, that is,
\begin{equation}\label{eq:init-cond}
\psi(x,0)=\psi^{(0)}(x)=\sqrt{1-x^2}\,\exp\left(\frac{i\,\alpha_{0}}{2\,\epsilon}\,x^{2}\right)\,,
\end{equation}
for different choices of the phase $\alpha_{0}$ and of the dispersion parameter $\epsilon\geq0$. In this case, the dispersionless solution (see table in Appendix \ref{sec:appendix}) will undergo asymptotic relaxation for $\alpha_{0}\geq2$, whereas for $\alpha_{0}<2$ it will exhibit a blow-up at finite time $t_{c}$ given by (\ref{eq:catatime}) and (\ref{eq:t-bu-phaseneg2}),
\begin{equation}\label{eq:catatime-num}
t_{c} =
\left\{
\begin{array}{lll}
\frac{\alpha_{0}}{4-\alpha_{0}^{2}} + \frac{4}{\left(4-\alpha_{0}^{2}\right)^{3/2}}\,\left[\frac{\pi}{2}+\arctan\left(\frac{\alpha_{0}}{\sqrt{4-\alpha_{0}^{2}}}\right)\right]
&&\mbox{if}\quad-2<\alpha_{0}<2\\
&&\\
\frac{1}{3}&&\mbox{if}\quad\alpha_{0}=-2\\
&&\\
-\frac{\alpha_{0}}{\alpha_{0}^{2}-4} + \frac{4}{\left(\alpha_{0}^{2}-4\right)^{3/2}}\,\,\mathrm{arctanh}\left(\frac{\sqrt{\alpha_{0}^{2}-4}}{\alpha_{0}}\right)
&&\mbox{if}\quad\alpha_{0}<-2\\
\end{array}
\right.
\qquad.
\end{equation}
As for the numerical integration of (\ref{eq:NLS_Foc}), the initial condition \eqref{eq:init-cond} features two gradient catastrophes at $x=\pm1$, hence, as such, it is not suitable for direct numerical implementation. We found it convenient to use the following infinite product representation of the square-root term in \eqref{eq:init-cond}:
$$
\prod_{j=1}^{\infty} e^{-\frac{x^{2j}}{2j}}=
\left\{
\begin{array}{lll}
\sqrt{1-x^{2}}&&-1\leq x\leq1\\
&&\\
0&&\mbox{else}
\end{array}\right.\,.
\label{eq:infinite-product}
$$
Using the above formula, we introduce the following smooth approximations of the initial condition \eqref{eq:init-cond}:
\begin{equation}\label{eq:init-cond-approx}
\psi^{(0)}_{N}(x)=\exp\left(\frac{i\,\alpha_{0}}{2\,\epsilon}\,x^{2}\right)\,\exp\left(-\sum_{j=1}^{N}\frac{x^{2j}}{2j}\right)\,.
\end{equation}
We integrate the focusing NLS equation \eqref{eq:NLS_Foc} with initial condition \eqref{eq:init-cond-approx} employing a step-adaptive pseudo-spectral method. While the space derivatives are evaluated using Fast Fourier Transform routines (FFTW) (\url{http://www.fftw.org/})
on a numerical box having size $L=64$ that is discretised with $2^{17}$ points, the time evolution is carried out using the numerical solver included in the ’ROCK4’ C++ library \cite{Abdulle2002}. The code relies on a fourth-order embedded-Runge-Kutta method suitable for large stiff problems. Moreover, the solver adapts each propagation time step to minimise the numerical error and control the scheme stability. We ensured that, at each time, the power spectrum for the spatial Fourier transforms does not reach the boundaries of the spectral domain. For the purposes of the present numerical investigation, we have observed that setting $N=10$ in (\ref{eq:init-cond-approx}) is sufficient to ensure a reliable comparison with the analytical solution of the dispersionless counterpart (\ref{eq:symsol}), as a further increase of $N>10$ for a fixed spatial discretisation step only affects the maximum of the absolute value of $|\psi|$ and its position over the space-time integration domain within a relative error of $1\%$.

As for the solution of the dispersionless counterpart, system (\ref{eq:ellsys-fluid}) with (\ref{eq:ini-Talanov-hydro}), we refer to it as semi-analytical, for it is obtained by directly numerically integrating (\ref{eq:3ODE-sym}) with (\ref{eq:init-3ODE-sym}) in time via a simple fourth-order embedded-Runge-Kutta method with relative accuracy of $10^{-9}$.

\begin{figure}[!h]
\begin{subfigure}{\textwidth}
  \centering
  \includegraphics[width=.8\linewidth]{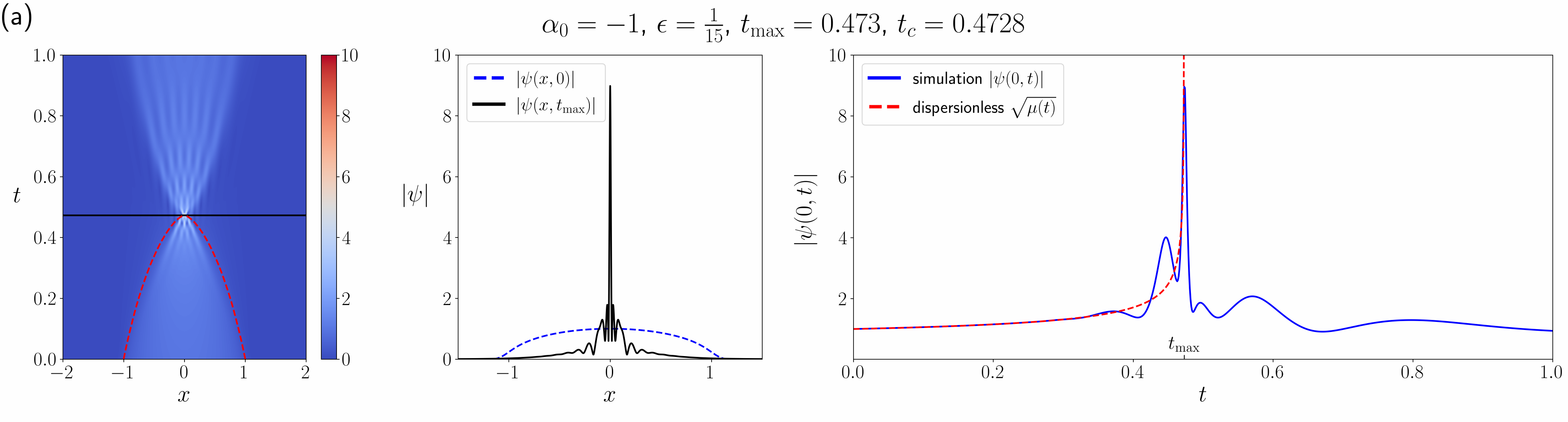}
\end{subfigure}\\\qquad\\
\begin{subfigure}{\textwidth}
  \centering
  \includegraphics[width=.8\linewidth]{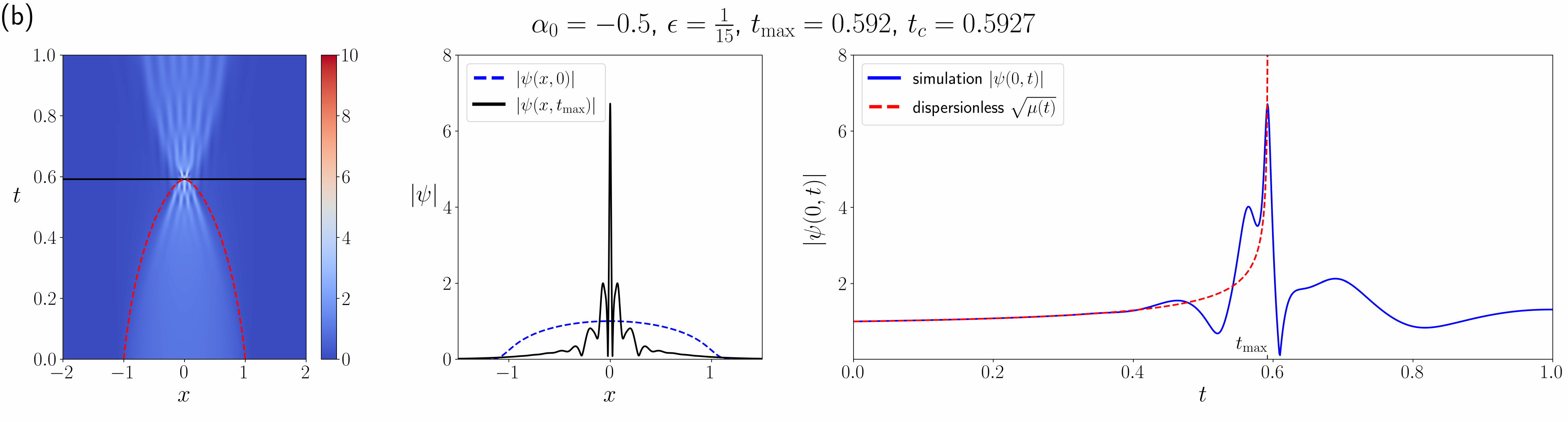}
\end{subfigure}\\\qquad\\
\begin{subfigure}{\textwidth}
  \centering
  \includegraphics[width=.8\linewidth]{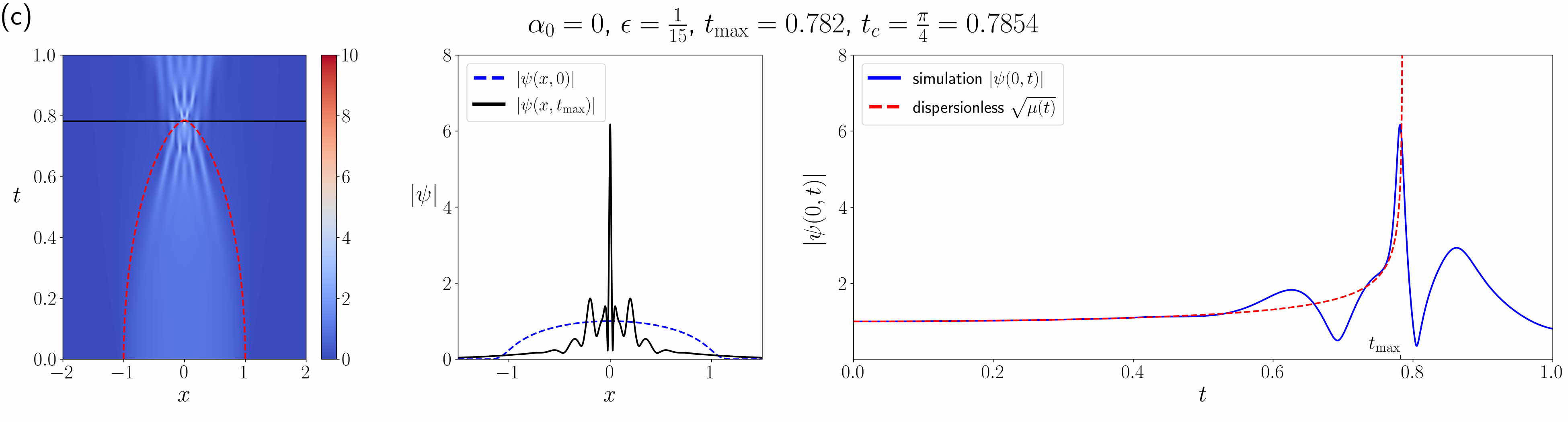}
\end{subfigure}\\\quad\\
\begin{subfigure}{\textwidth}
  \centering
  \includegraphics[width=.8\linewidth]{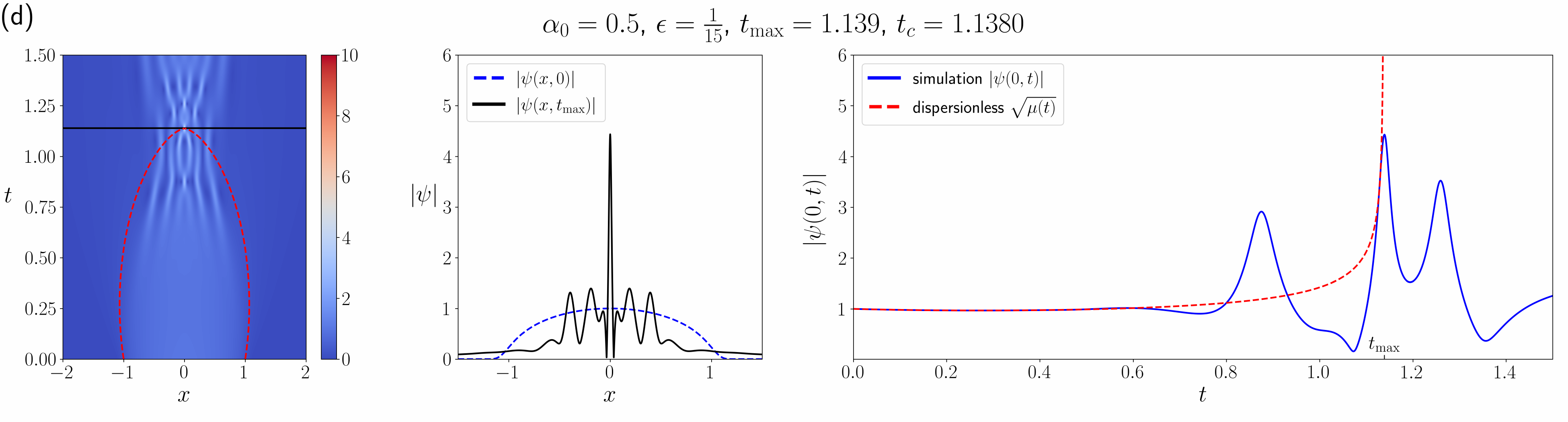}
\end{subfigure}
\caption{Numerical evolution of the initial condition $\psi^{(0)}_{10}(x)$ given by (\ref{eq:init-cond-approx}) for $\epsilon=\frac{1}{15}$ and for various choices of $\alpha_{0}<2$, where the dispersionless solution features a blow-up at fine time $t_{c}$. In each group of figures, the figure on the left shows the evolution of the absolute value of $\psi$ as an $(x,t)$-plot, where the solid black line indicates the time of the maximum $t_{\mathrm{max}}$ and the dashed red line indicates the endpoints of the support of the corresponding dispersionless solution (\ref{eq:support}); the figure in the centre shows the initial condition as a dashed blue line and the absolute value of $\psi$ at $t=t_{\mathrm{max}}$ as a solid black line; the figure on the right shows the numerical evolution of the absolute value of $\psi$ at $x=0$ for all values of $t$ in the range of numerical integration as a solid blue line, overlapping the analytical evolution of the corresponding dispersionless solution at $x=0$ (namely, $\sqrt{\mu(t)}$, see (\ref{eq:curvmaxpar})) as a dashed red line. It is possible to observe that, as the dispersionless semi-analytical solution features a blow-up at fine time $t_{c}$, the corresponding dispersive numerical solution features a local maximum at time $t_{\mathrm{max}}$, where $t_{c}$ and $t_{\mathrm{max}}$ are in excellent agreement.}
\label{Fig:plot_10G_eps0067_maxima}
\end{figure}

\begin{figure}[!h]
\begin{subfigure}{\textwidth}
  \centering
  \includegraphics[width=.8\linewidth]{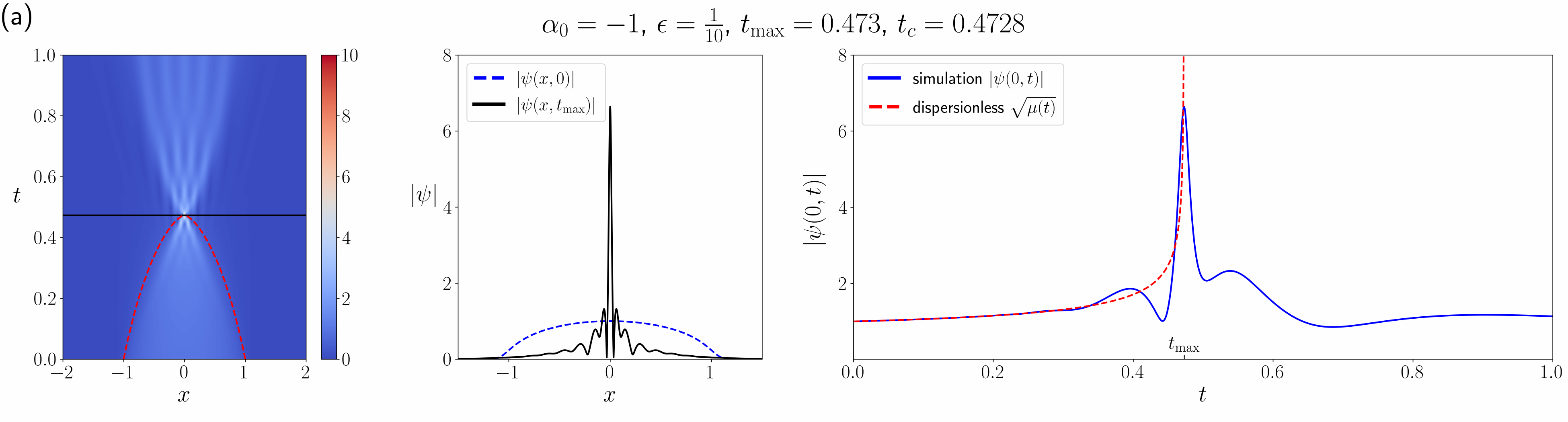}
\end{subfigure}\\\qquad\\
\begin{subfigure}{\textwidth}
  \centering
  \includegraphics[width=.8\linewidth]{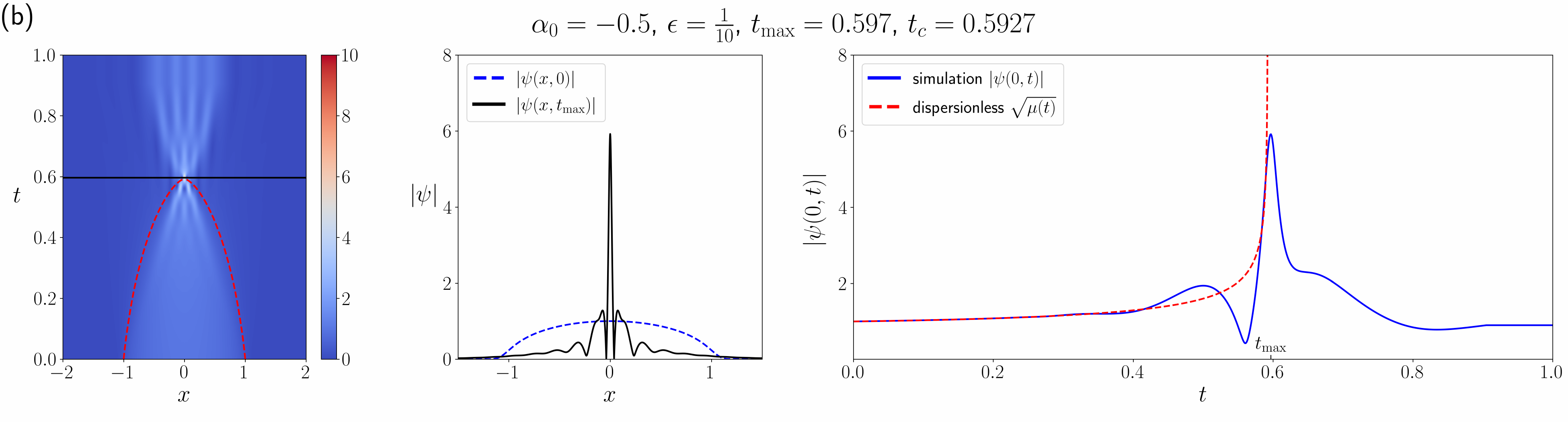}
\end{subfigure}\\\qquad\\
\begin{subfigure}{\textwidth}
  \centering
  \includegraphics[width=.8\linewidth]{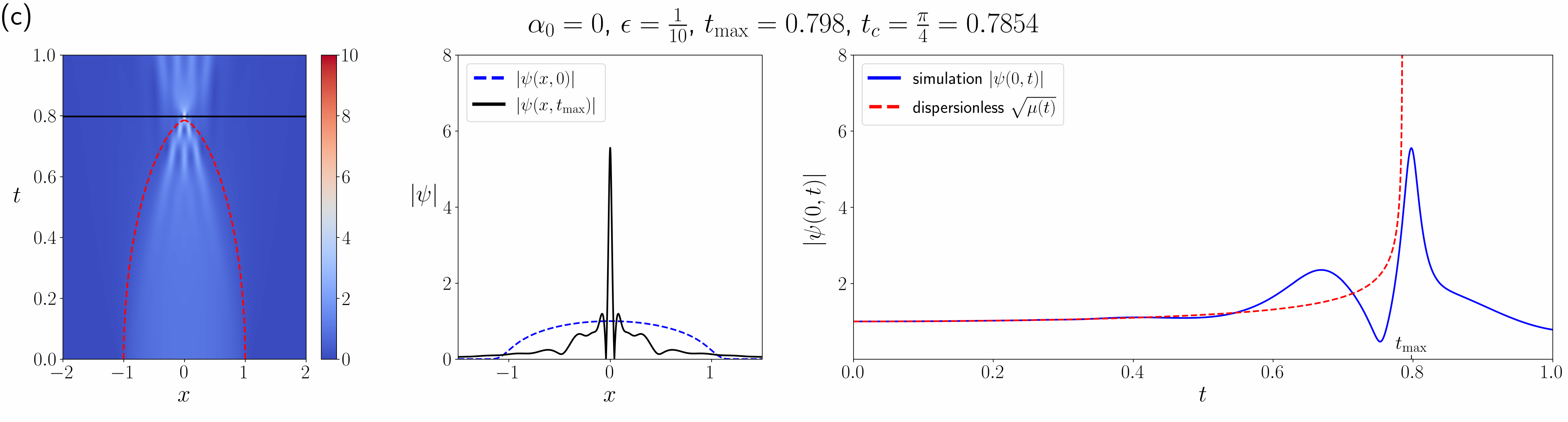}
\end{subfigure}\\\quad\\
\begin{subfigure}{\textwidth}
  \centering
  \includegraphics[width=.8\linewidth]{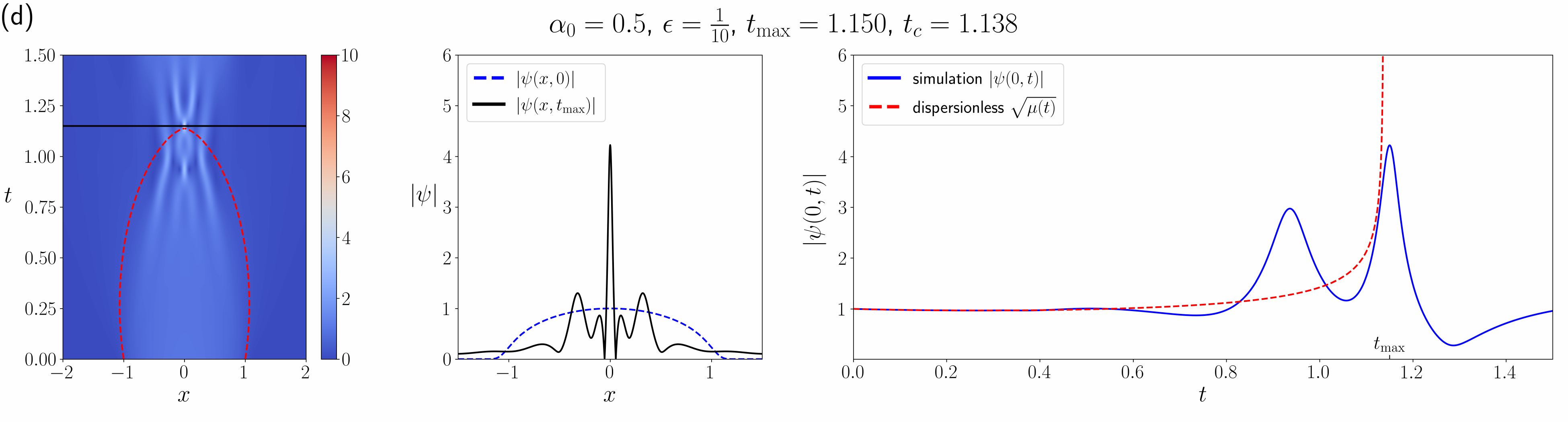}
\end{subfigure}
\caption{Numerical evolution of the initial condition $\psi^{(0)}_{10}(x)$ given by (\ref{eq:init-cond-approx}) for $\epsilon=\frac{1}{10}$ and for various choices of $\alpha_{0}<2$, where the dispersionless solution features a blow-up at fine time $t_{c}$. In each group of figures, the figure on the left shows the evolution of the absolute value of $\psi$ as an $(x,t)$-plot, where the solid black line indicates the time of the maximum $t_{\mathrm{max}}$ and the dashed red line indicates the endpoints of the support of the corresponding dispersionless solution (\ref{eq:support}); the figure in the centre shows the initial condition as a dashed blue line and the absolute value of $\psi$ at $t=t_{\mathrm{max}}$ as a solid black line; the figure on the right shows the numerical evolution of the absolute value of $\psi$ at $x=0$ for all values of $t$ in the range of numerical integration as a solid blue line, overlapping the analytical evolution of the corresponding dispersionless solution at $x=0$ (namely, $\sqrt{\mu(t)}$, see (\ref{eq:curvmaxpar})) as a dashed red line. It is possible to observe that, as the dispersionless semi-analytical solution features a blow-up at fine time $t_{c}$, the corresponding dispersive numerical solution features a local maximum at time $t_{\mathrm{max}}$, where $t_{c}$ and $t_{\mathrm{max}}$ are in excellent agreement.}
\label{Fig:plot_10G_eps01_maxima}
\end{figure}

\begin{figure}[!h]
\centering
\begin{subfigure}{0.21\textwidth}
  \centering
  \includegraphics[width=\linewidth]{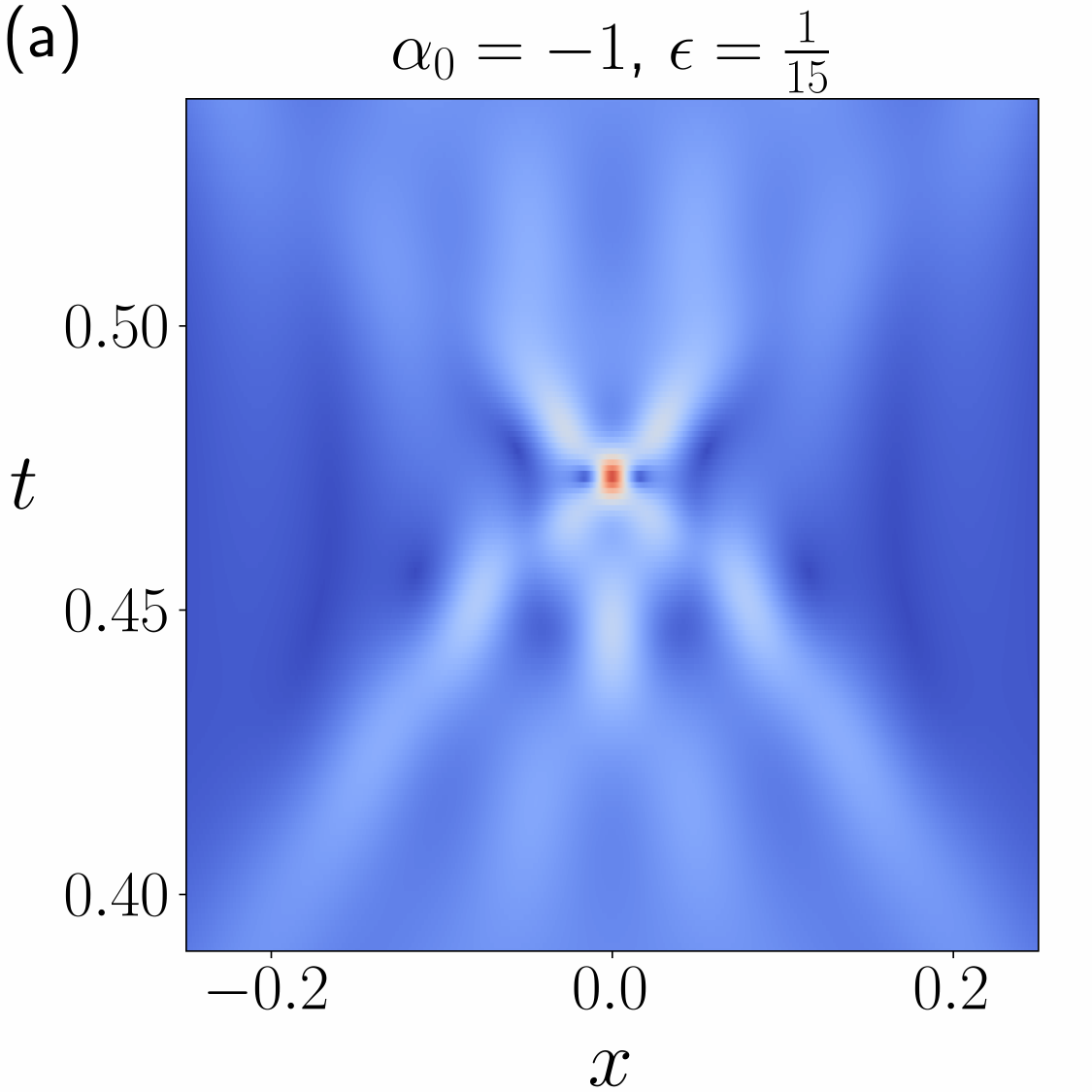}
\end{subfigure}\quad
\begin{subfigure}{0.21\textwidth}
  \centering
  \includegraphics[width=\linewidth]{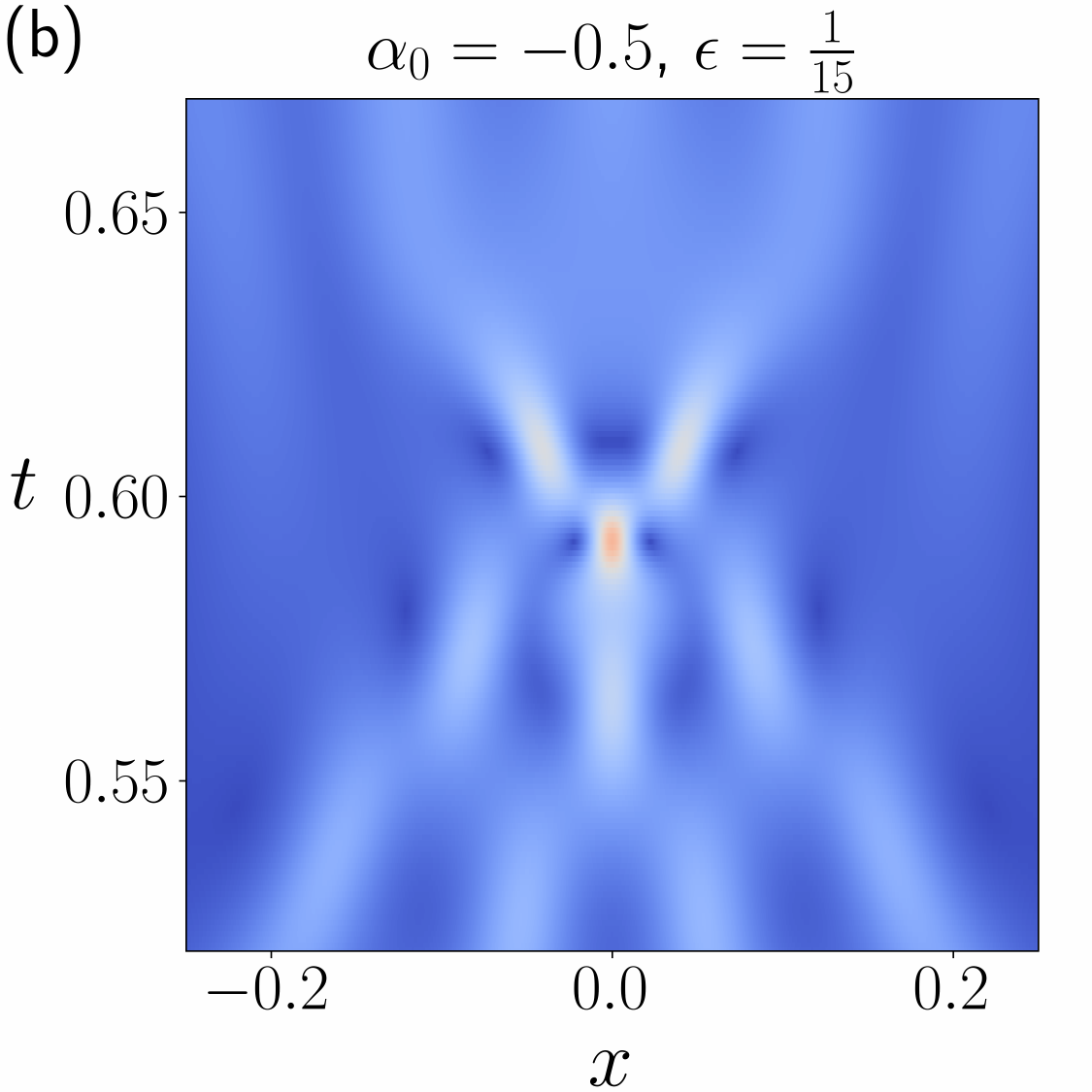}
\end{subfigure}\quad
\begin{subfigure}{0.21\textwidth}
  \centering
  \includegraphics[width=\linewidth]{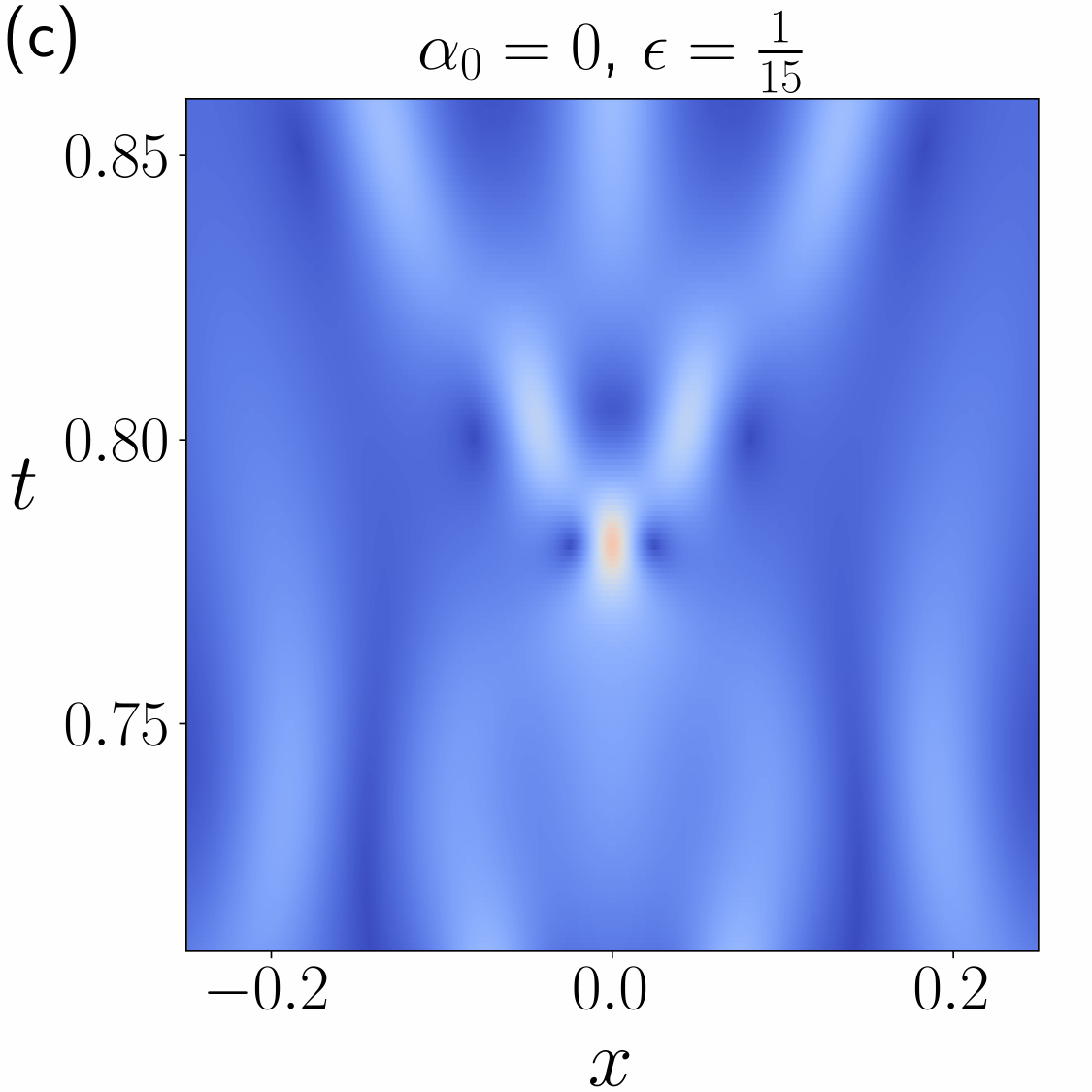}
\end{subfigure}\quad
\begin{subfigure}{0.21\textwidth}
  \centering
  \includegraphics[width=\linewidth]{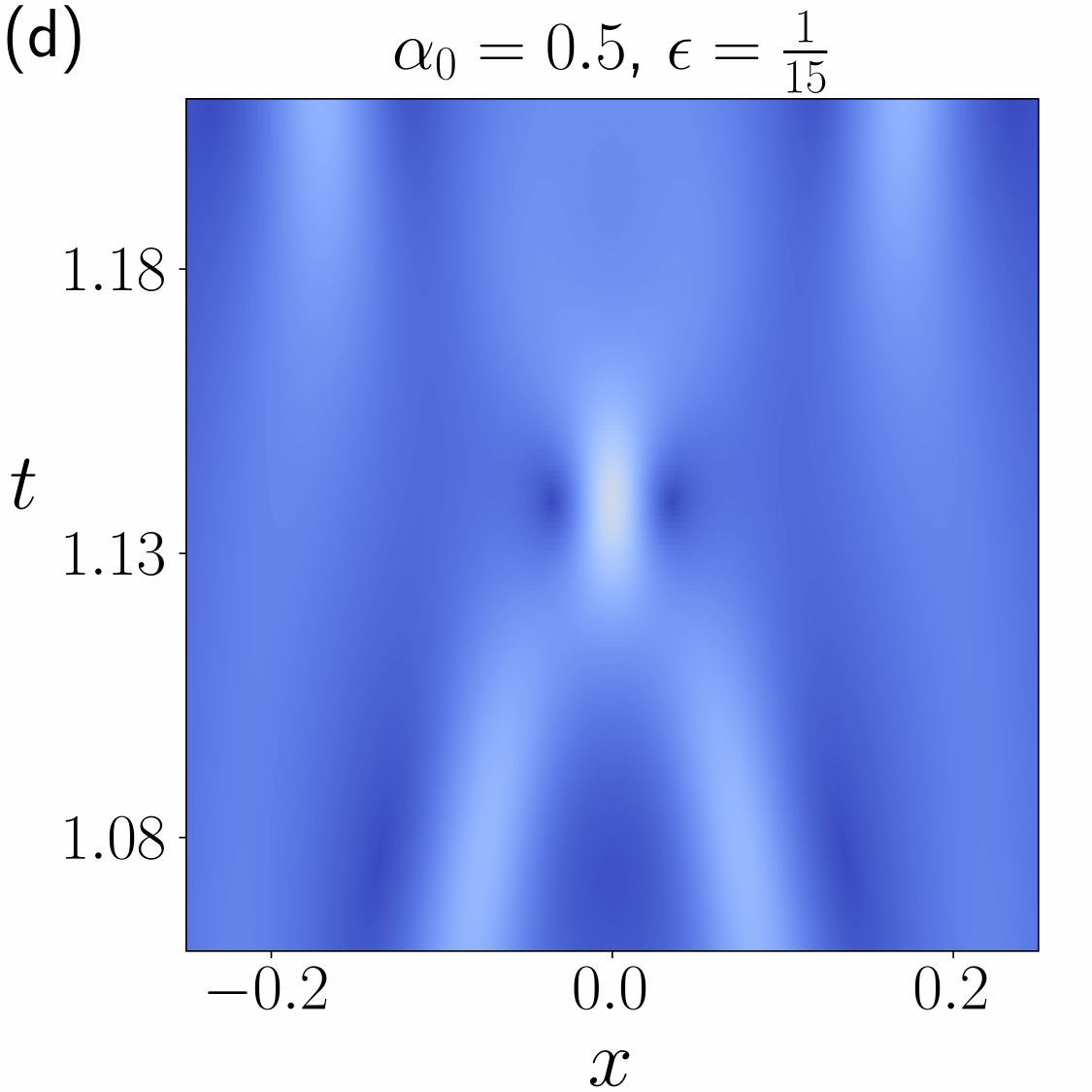}
\end{subfigure}
\caption{Magnification of the $(x,t)$-plot around the peak for each one of the cases in Figure \ref{Fig:plot_10G_eps0067_maxima} ($\epsilon=\frac{1}{15}$), showing a rogue-wave like behaviour, with two troughs neighbouring the maximum, as in the Peregrine soliton. Subfigures a-d refer to the corresponding subfigures a-d of Figure \ref{Fig:plot_10G_eps0067_maxima}, with the colour scale remaining the same.}
\label{Fig:plot_10G_eps0067_maxima_zooms}
\end{figure}

\begin{figure}[!h]
\centering
\begin{subfigure}{0.21\textwidth}
  \centering
  \includegraphics[width=\linewidth]{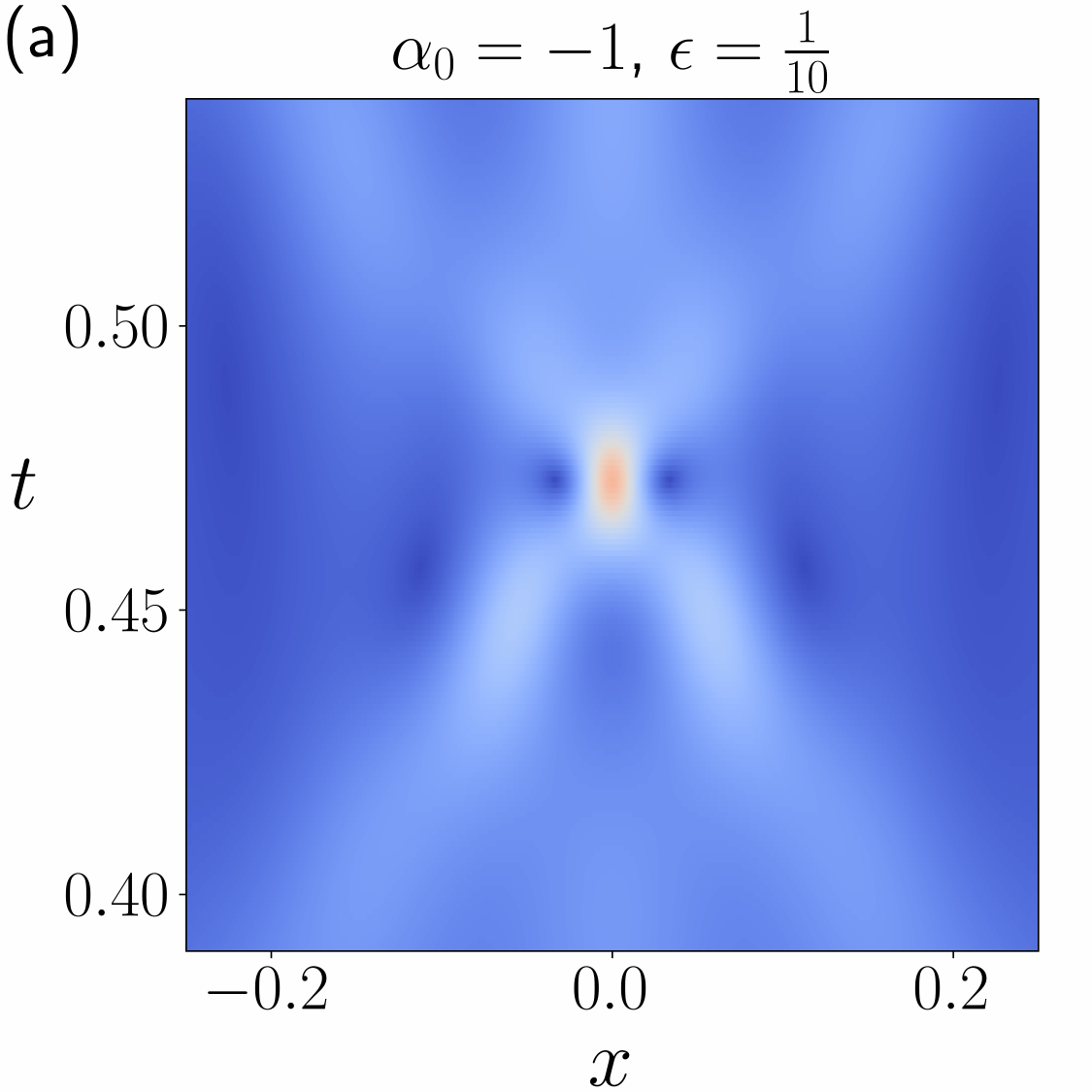}
\end{subfigure}\quad
\begin{subfigure}{0.21\textwidth}
  \centering
    \includegraphics[width=\linewidth]{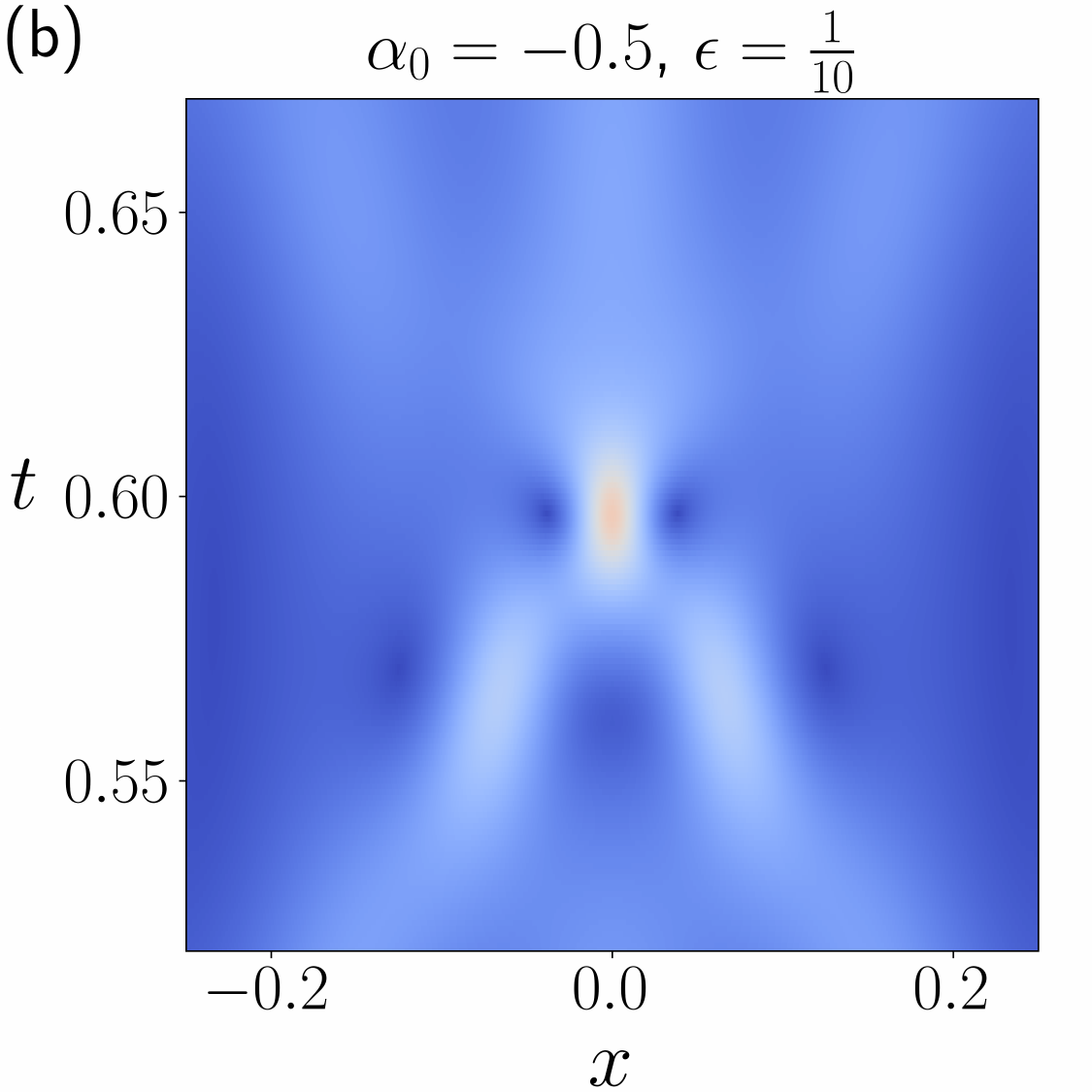}
\end{subfigure}\quad
\begin{subfigure}{0.21\textwidth}
  \centering
  \includegraphics[width=\linewidth]{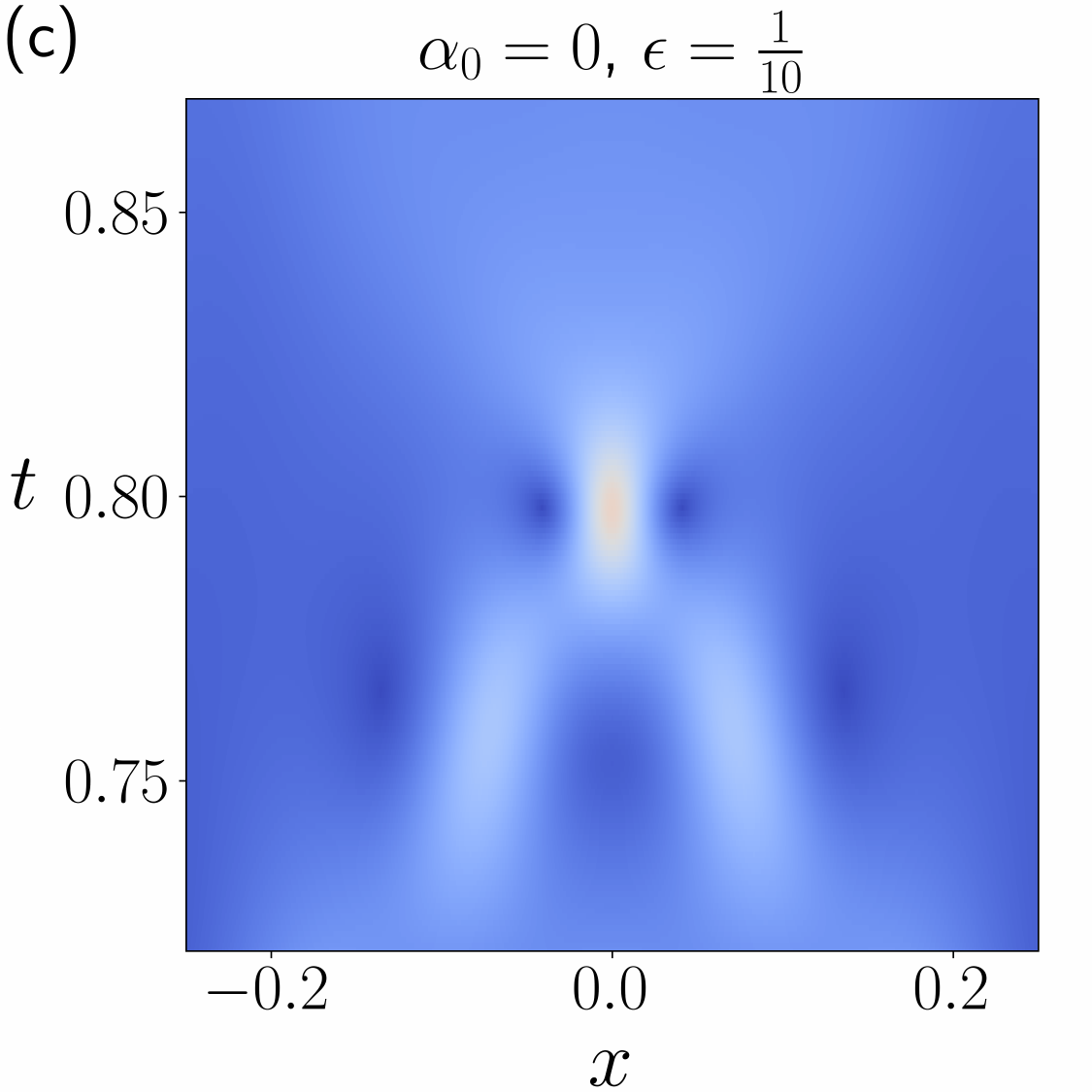}
\end{subfigure}\quad
\begin{subfigure}{0.21\textwidth}
  \centering
  \includegraphics[width=\linewidth]{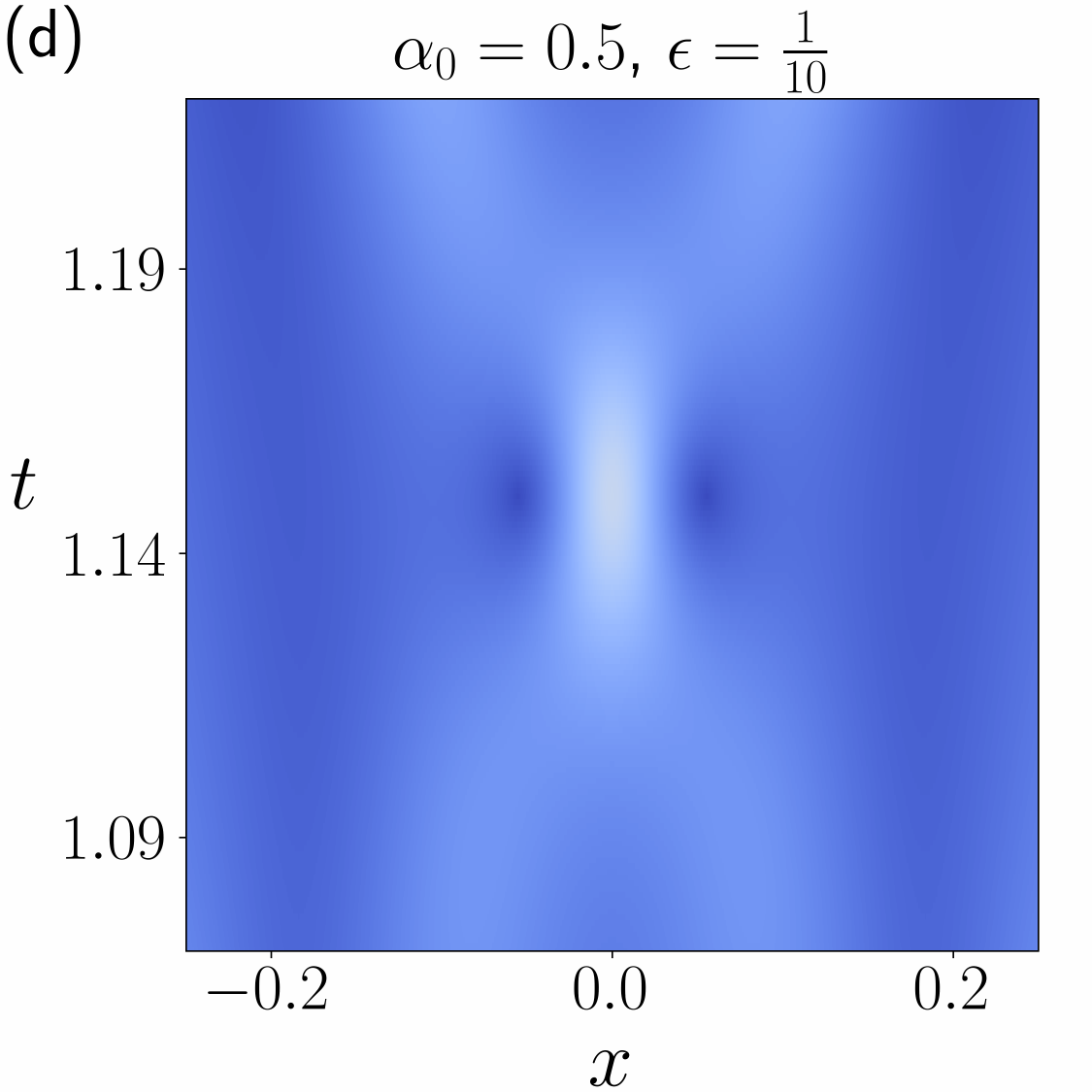}
\end{subfigure}
\caption{Magnification of the $(x,t)$-plot around the peak for each one of the cases in Figure \ref{Fig:plot_10G_eps01_maxima} ($\epsilon=\frac{1}{10}$), showing a rogue-wave like behaviour, with two troughs neighbouring the maximum, as in the Peregrine soliton. Subfigures a-d refer to the corresponding subfigures a-d of Figure \ref{Fig:plot_10G_eps01_maxima}, with the colour scale remaining the same.}
\label{Fig:plot_10G_eps01_maxima_zooms}
\end{figure}

As illustrated in Figures \ref{Fig:plot_10G_eps0067_maxima}-\ref{Fig:plot_10G_2phase_decay}, the numerical solution of the NLS equation appears to be driven by the behaviour of the corresponding dispersionless counterpart. For $\alpha_0< 2$, where the dispersionless solution evolving from (\ref{eq:ini-Talanov-hydro}) is expected to blow-up at finite time $t_{c}$ (\ref{eq:catatime-num}), the dispersive NLS solution features a local maximum for $|\psi|$ at a time $t_{\mathrm{max}}$ which is remarkably well approximated by $t_{c}$, $t_{\mathrm{max}}\approx t_{c}$ (see Figure \ref{Fig:plot_tc_vs_alpha0}).  Due to its similarity to the Peregrine soliton (see Figures \ref{Fig:plot_10G_eps0067_maxima_zooms}a-d and \ref{Fig:plot_10G_eps01_maxima_zooms}a-d) and its height of up to 10 times the initial height of the parabola $\rho(x,0)$ (for instance, see Figure \ref{Fig:plot_10G_eps0067_maxima}a), we suggest an interpretation of the peak as a rogue-wave solution. The process leading to the formation of these structure qualitatively resembles the Bertola-Tovbis  regularisation mechanism of the gradient catastrophe through the generation of a Peregrine breather \cite{BertolaTovbis2013,TikanRobertiElRandouxSuret2022}. However, here the presence of additional lobes on the side of the main peak suggests the presence of a high-order rogue wave \cite{HeZhangWangPorsezianFokas2013}.
 In particular for $\alpha_0 \leq 0$ (Figures \ref{Fig:plot_10G_eps0067_maxima}a-c and \ref{Fig:plot_10G_eps01_maxima}a-c), the focusing behaviour  and the rogue wave formation are strengthened by the negative chirp and dominate the dynamics.    Whereas, for  $0<\alpha_0<2$ (Figures \ref{Fig:plot_10G_eps0067_maxima}d and \ref{Fig:plot_10G_eps01_maxima}d) the focusing process is increasingly  impeded  by the positive phase and in competition with the  oscillations propagating towards the vertex of the parabola from the two vacuum points at $x=\pm1$. These oscillations are numerically reminiscent of the dam break flow observed and studied in \cite{ElKhamisTovbis2016}, suggesting a similar underlying mechanism. However, it is worth observing that the box-like initial condition for the dam break problem in \cite{ElKhamisTovbis2016} has two symmetric discontinuity points in the function at the endpoints of the compact support but not in the first spatial derivative, whereas the initial condition (\ref{eq:ini-Talanov}) is continuous but features a divergence on the first derivative at the two vacuum points coinciding with the endpoints of the compact support. Lastly, for $\alpha_0\geq 2$ -- see Figures \ref{Fig:plot_10G_2phase_decay} -- where the dispersionless solution evolving from (\ref{eq:ini-Talanov-hydro}) is expected to relax, the NLS solution behaves accordingly for all the values of the dispersion parameter $\epsilon$ for which it was possible to carry out the numerical integration (the largest of such values being $\epsilon=1$, well beyond the semiclassical limit).

The numerical evidence, supporting the analytical predictions from the dispersionless model, suggests that the rogue wave formation (or suppression) for the NLS equation  \eqref{eq:NLS_Foc} with  \eqref{eq:init-cond-approx} can be controlled and manipulated by conveniently tuning the chirp and the amplitude of the initial parabolic profile, analogously to what is observed and tested in \cite{TikanRobertiElRandouxSuret2022} for a sech profile.

\renewcommand\thesubfigure{\alph{subfigure}}
\addtocounter{figure}{0}
\begin{figure}[!h]
\centering
\includegraphics[width=0.6\linewidth]{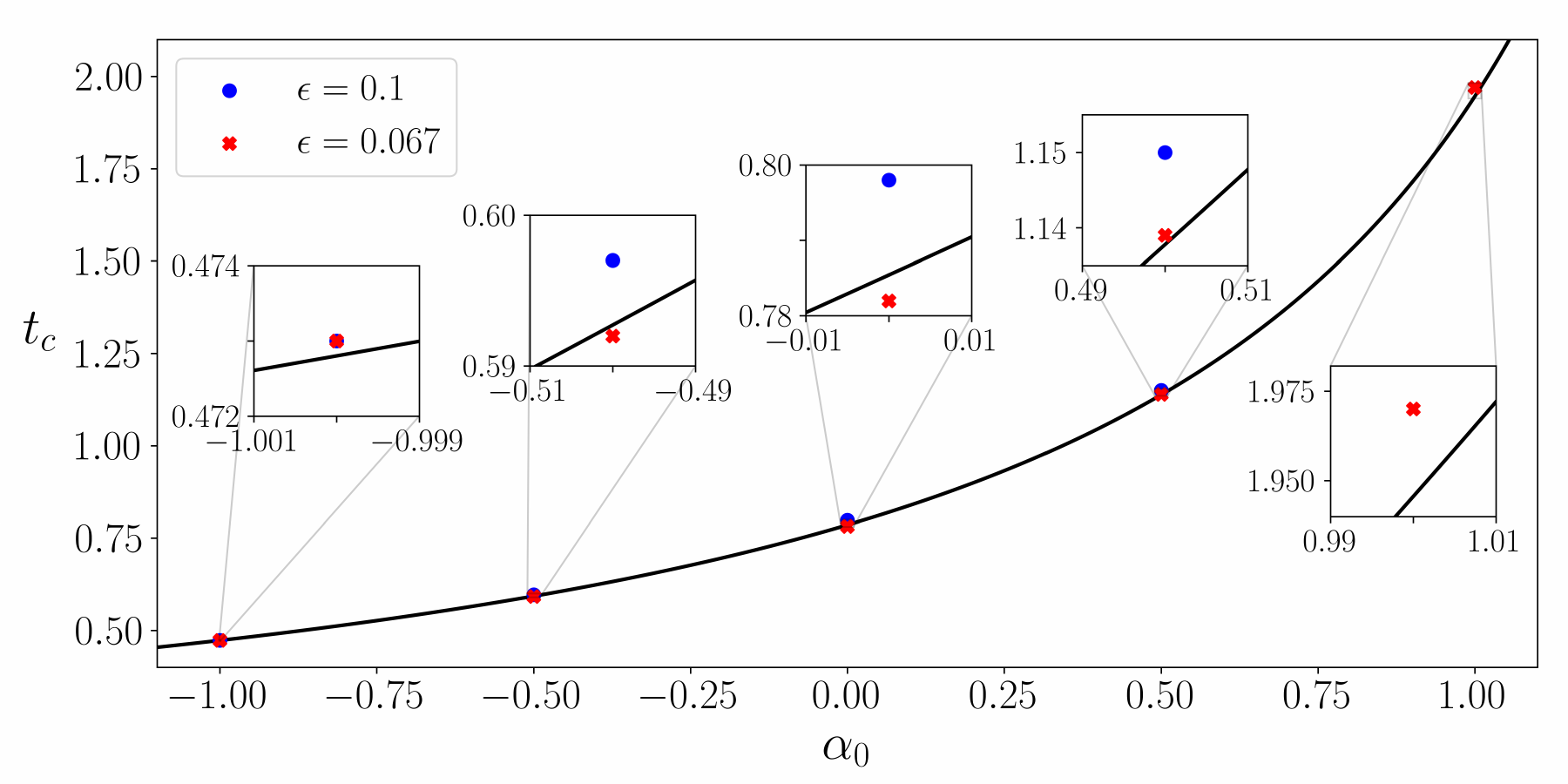}
\caption{The catastrophe time $t_{c}$ for the dispersionless solution emerging from the initial condition (\ref{eq:init-cond-approx}) as a function of the phase $\alpha_{0}<2$ (solid line), drawn along with the values of $t_{\mathrm{max}}$ against the corresponding values of $\alpha_{0}$ from the numerical experiments (blue dots for $\epsilon=\frac{1}{10}$ and red crosses for $\epsilon=\frac{1}{15}$). The time of the maximum $t_{\mathrm{max}}$ is remarkably well approximated by the time of the catastrophe $t_{c}$. For $\alpha_0=1$ and $\epsilon=0.1$, where we do not have a blue point in the plot, the oscillations propagating from the vacuum points at $x=\pm1$ towards the peak of the parabola in the time evolution eventually become comparable with the peak that forms under the drive of the dispersionless solution. This phenomenon might be amplified by numerical instabilities.}
\label{Fig:plot_tc_vs_alpha0}
\end{figure}

\begin{figure}[!h]
\begin{subfigure}{\textwidth}
  \centering
  \includegraphics[width=.8\linewidth]{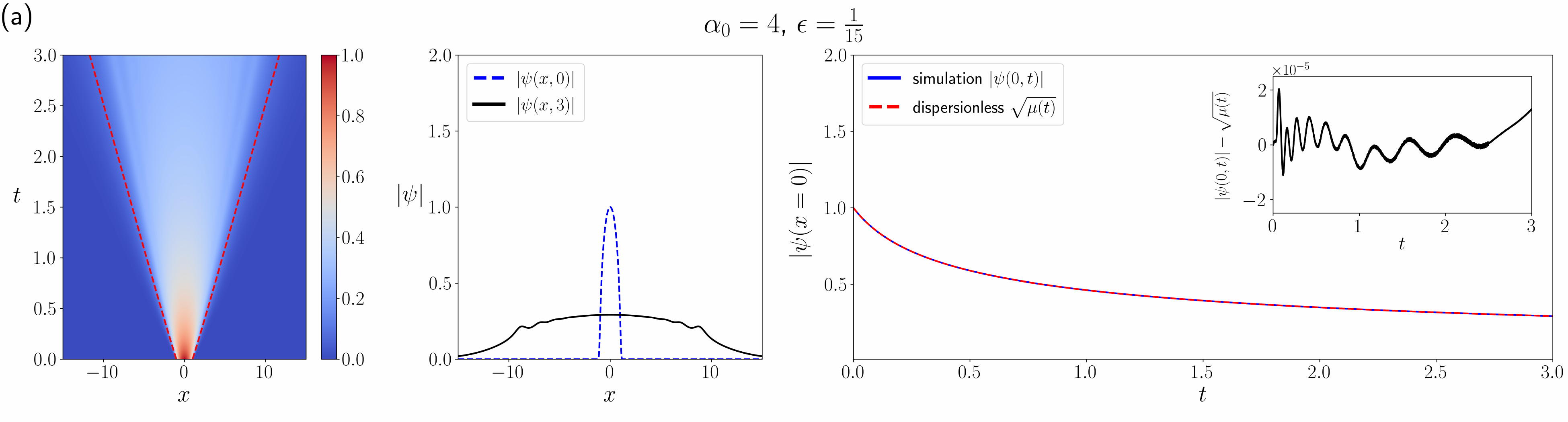}
\end{subfigure}\\\quad\\
\begin{subfigure}{\textwidth}
  \centering
  \includegraphics[width=.8\linewidth]{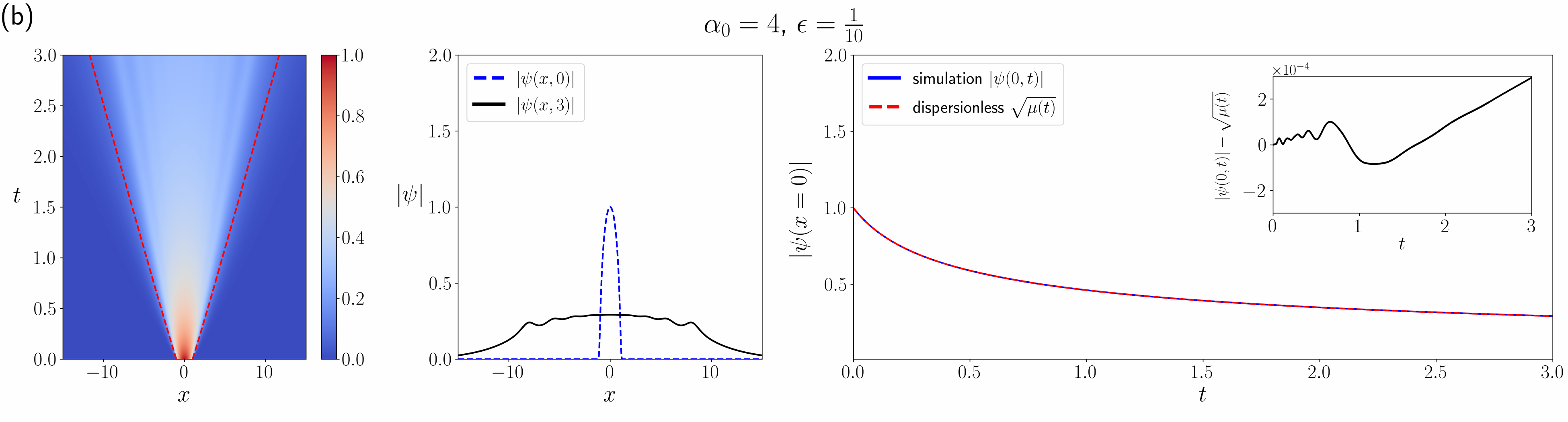}
\end{subfigure}\\\quad\\
\begin{subfigure}{\textwidth}
  \centering
  \includegraphics[width=.8\linewidth]{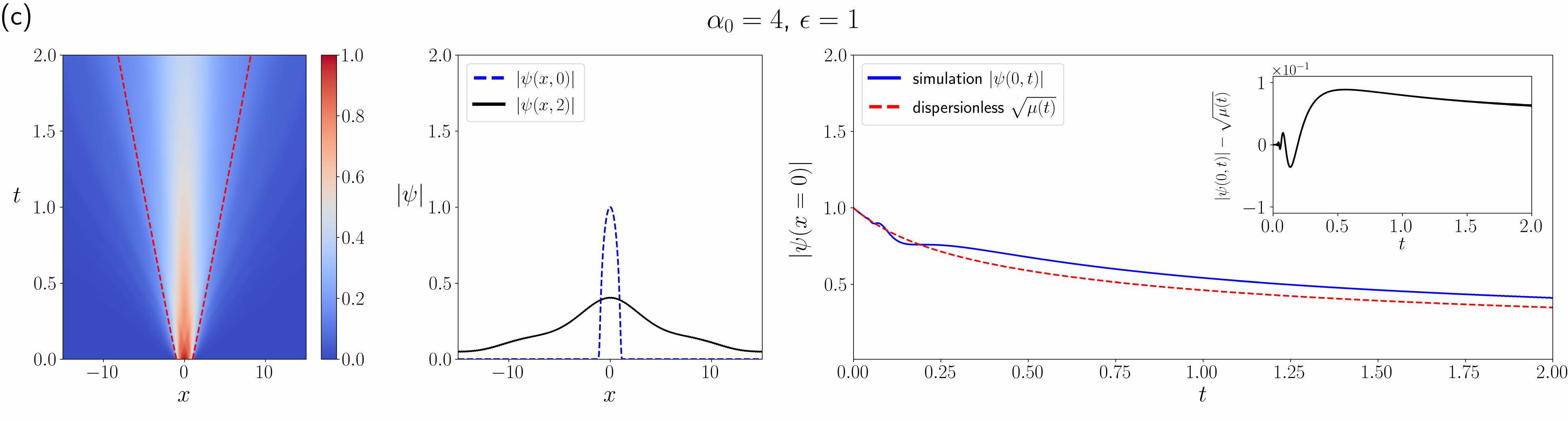}
\end{subfigure}
\caption{Numerical evolution of the initial condition $\psi^{(0)}_{10}(x)$ given by (\ref{eq:init-cond-approx}) for different values of $\epsilon$ and for $\alpha_{0}=4>2$, where the dispersionless solution undergoes asymptotic relaxation. In each group of figures, the figure on the left shows the evolution of the absolute value of $\psi$ as an $(x,t)$-plot, where the dashed red line indicates the endpoints of the support of the corresponding dispersionless solution (\ref{eq:support}); the figure in the centre shows the initial condition as a dashed blue line and the absolute value of $\psi$  at the end of the time integration interval 
as a solid black line; the figure on the right shows the numerical evolution of the absolute value of $\psi$ at $x=0$ for all values of $t$ in the range of numerical integration as a solid blue line, overlapping the analytical evolution of the corresponding dispersionless solution at $x=0$ (namely, $\sqrt{\mu(t)}$, see (\ref{eq:curvmaxpar})) as a dashed red line. It is possible to observe that in the small dispersion (semiclassical) regime, the dispersionless solution accurately approximates the dispersive solution. Remarkably, even for large $\epsilon$, well outside the semiclassical regime, it is possible to predict the relaxation behaviour of the dispersive solution based on the relaxation of the dispersionless counterpart.}
\label{Fig:plot_10G_2phase_decay}
\end{figure}

\section{Conclusion and future directions}

In this paper we numerically analysed a condition for the rogue wave formation for the focusing NLS equation (\ref{eq:NLS_Foc}) for a class of compactly supported initial data. Such criterion is based on the study of self-similar solutions of the elliptic dispersionless NLS system (\ref{eq:ellsys-fluid}).

A visual comparison between Figures \ref{Fig:plot_10G_eps0067_maxima}c and \ref{Fig:plot_10G_eps01_maxima}c shows that the maximum elevation of the central peak increases when the dispersion decreases: this qualitative observation suggests that -- at least partly -- it should be possible to describe the behaviour of such peak in terms of the solution of the purely dispersionless system.

The investigation of the blow-up of parabolic solutions of dispersionless NLS models was initiated in the paper \cite{GurevichShvartsburg1970} and, in its extended version, in the book \cite{GurevichShvartsburg1973} (which, to the best of our knowledge, to date has never been fully translated into English), in the context of studying radiowave beams of different profiles interacting with the ionosphere: there the authors found the criterion for the blow-up of a parabolic profile in the absence of dispersion for a subclass of the initial data analysed in the present paper (adding some considerations about perturbing such profile).

Here we complete the classification and characterisation of the compactly supported self-similar solutions of dispersionless focusing NLS equation and we perform an accurate numerical comparison with the dispersive counterpart. This numerical study confirms that the dispersionless blow-up criterion is a surprisingly good indicator of the rogue wave formation in the dispersive model. We also remark that such criterion qualitatively holds for rather large values of the dispersive parameters ($\epsilon = \frac{1}{15}$ and $\epsilon = \frac{1}{10}$), well outside the semi-classical approximation regime.

The compact support of the initial conditions is another peculiarity of the paper. In the case of smooth initial data, a thorough investigation is presented in \cite{BertolaTovbis2013}, where the rogue wave formation is studied also from the analytical point of view. Here the presence of vacuum sectors in the initial data yields non-trivial numerical and analytical complications, but the self-similarity of the dispersionless evolution allows to identify anyway a criterion for the formation of rogue waves.

Moreover, we numerically observe that the evolution leading to the formation of the rogue wave shows remarkable similarities with the scenarios presented in \cite{BertolaTovbis2013} and \cite{ElKhamisTovbis2016}, the prevalence of one or the other seeming to depend on the sign of a function of the chirp and of the amplitude. As chirp and amplitude are easy to control in water tank wave generators (see \cite{TikanRobertiElRandouxSuret2022}) and in nonlinear optical devices, we suggest that our result may be relevant to and put to test in future experiments in optics and fluid dynamics.

We finally remark that the parabolic structure of self-similar solutions appears to be independent on the number of spatial dimensions. This can be verified by direct inspection.
Actually, it is well known \cite{Madelung1927} that the focusing NLS equation with function $\psi=\sqrt{\rho} \exp\left({\frac{i}{\epsilon} \theta} \right)$ can be mapped through the Madelung transformation in a fluid-dynamic-like system, whose dispersionless counterpart is the potential shallow water equation
\begin{equation}
\mathbf{u}_t+\mathbf{u} \cdot \nabla \mathbf{u} -\nabla \rho=0,  \qquad
\rho_t +\nabla \cdot (\rho \mathbf{u} )=0\, , \qquad \mathbf{u} = \nabla \theta.
\end{equation}
Such system admits, as in the one-dimensional case, a class of quadratic exact solutions (see \cite{Thacker1981} for hyperbolic case appearing in fluid dynamics context)
\begin{equation}
\theta=\theta_{20}(t) \frac{x^2}{2}+\ic(t) x y+ \theta_{02}(t) \frac{y^2}{2}\, , \qquad
\rho=\rho_{20}(t) x^2+\rho_{02}(t) y^2 +\rho_{11}(t)xy+ \rho_{00}(t)\, ,
\end{equation}
whose coefficients satisfy the ODE system generalizing (\ref{eq:3ODE-sym})
\begin{equation}
\begin{split}
&\dot{\theta}_{20}+\theta_{20}^2+\theta_{11}^2  -2 \rho_{20}=0 \, , \qquad
\dot{\theta}_{11}+ \ic(\theta_{20}+\theta_{02}) - \rho_{11}=0\, , \qquad
 \dot{\theta}_{02}+\theta_{02}^2+\ic^2 -2 \rho_{02}=0 \, , \\
& \dot{\rho}_{20}+ \rho_{20} (3\theta_{20}+\theta_{02}) +\rho_{11} \ic=0\, , \qquad
\dot{\rho}_{02}+  \rho_{02} (\theta_{20}+3\theta_{02}) +\rho_{11} \ic =0\, , \\
&\dot{\rho}_{11}+ 2\rho_{11}(\theta_{20}+\theta_{02}) +2 \ic (\rho_{20} + \rho_{02})  =0\, , \qquad
\dot{\rho} _{00}+\rho _{00} ( \theta_{20}+\theta_{02})=0 \, .
\end{split}
\label{quad-sys}
\end{equation}
An interesting direction for future investigation is the multidimensional extension of the analysis herein presented, although this is expected to be a much more delicate study
(as many new behaviors seem to appear).

\appendix
\begin{landscape}
\section{Appendix: Classification of self-similar solutions for negative $\boldsymbol{\gamma_0}$}
\label{sec:appendix}
\begin{table}[h]
\centering
$$
\begin{array}{c|c|c}
\text{phase parameter } \alpha_0 & \sigma \text{ range}& \text{Solution } t=t(\sigma)\\
\hline && \\
\alpha_0> 2 \sqrt{-\gamma_0} & 1>\sigma>0&t =
 \frac{\sqrt{A+B \sigma}}{A\sigma }
  -\frac{\sqrt{A+B }}{A }
 +
\frac{B}{2 A^{3/2}} \log \left( \frac{(\sqrt{A+B}+\sqrt{A})(\sqrt{A+B \sigma}-\sqrt{A})}{(\sqrt{A+B}-\sqrt{A})(\sqrt{A+B \sigma}+\sqrt{A})}\right) \\
&& \\ \hline&& \\
\alpha_0= 2 \sqrt{-\gamma_0} & 1>\sigma>0& t=\frac{1-\sigma ^{3/2}}{3 \sqrt{-\gamma _0} \sigma ^{3/2}} \\
&& \\ \hline&& \\
 \begin{array}{c}
0<\alpha_0< 2 \sqrt{-\gamma_0}\\  (\text{nonmonotonic solution})
\end{array}
& \begin{array}{c}
1>\sigma>1-\frac{\alpha_0^2}{4|\gamma_0|}
\\ \\
\sigma>1-\frac{\alpha_0^2}{4|\gamma_0|}
\end{array}&
\begin{array}{c}
t= \frac{\sqrt{A+B \sigma}}{A \sigma }-
\frac{B\, }{(-A)^{3/2}}  \mathrm{arctan} \left(\frac{\sqrt{A+B \sigma}}{\sqrt{-A}}\right)
-\frac{\sqrt{A+B }}{A }+
\frac{B\, }{(-A)^{3/2}}  \mathrm{arctan} \left(\frac{\sqrt{A+B }}{\sqrt{-A}}\right) \\ \\
t= -\frac{\sqrt{A+B \sigma}}{A \sigma }+
\frac{B\, }{(-A)^{3/2}}  \mathrm{arctan} \left(\frac{\sqrt{A+B \sigma}}{\sqrt{-A}}\right)
-\frac{\sqrt{A+B }}{A }+
\frac{B\, }{(-A)^{3/2}}  \mathrm{arctan} \left(\frac{\sqrt{A+B }}{\sqrt{-A}}\right)
\end{array}
\\
&& \\ \hline&& \\
\begin{array}{c}
\alpha_0=0 \\
\text{ (see \cite{GurevichShvartsburg1970})}
\end{array}
& \sigma>1 &t=\frac{\sqrt{\sigma -1}}{2 \sqrt{-\gamma _0} \sigma }+\frac{1}{2 \sqrt{-\gamma _0}} \mathrm{arctan}\left(\sqrt{\sigma -1}\right)\\
&& \\ \hline&& \\
-2 \sqrt{-\gamma_0}<\alpha_0<0  & \sigma>1& t=- \frac{\sqrt{A+B \sigma}}{A \sigma }+
\frac{B\, }{(-A)^{3/2}}  \mathrm{arctan} \left(\frac{\sqrt{A+B \sigma}}{\sqrt{-A}}\right)
+\frac{\sqrt{A+B }}{A }-
\frac{B\, }{(-A)^{3/2}}  \mathrm{arctan} \left(\frac{\sqrt{A+B }}{\sqrt{-A}}\right) \\
&& \\ \hline&& \\
\alpha_0= -2 \sqrt{-\gamma_0} & \sigma>1& t=\frac{1-{\sigma ^{-3/2}}}{3 \sqrt{-\gamma_0}}\\
&& \\ \hline&& \\
\alpha_0< -2 \sqrt{-\gamma_0} & \sigma>1&  t= -\frac{\sqrt{A+B \sigma}}{A\sigma }
  +\frac{\sqrt{A+B }}{A }
  -\frac{B}{2 A^{3/2}} \log \left( \frac{(\sqrt{A+B}+\sqrt{A})(\sqrt{A+B \sigma}-\sqrt{A})}{(\sqrt{A+B}-\sqrt{A})(\sqrt{A+B \sigma}+\sqrt{A})}\right) \\
  && \\
\end{array}
$$
\caption{$A=\alpha_0^2+4\gamma_0$ and $B=-4 \gamma_0 >0$. The blow-up derivative happens when $\alpha_0 < 2 \sqrt{-\gamma_0}$.}
\end{table}
\end{landscape}

\subsection*{Acknowledgements}
The authors would like to thank the Mathematics of Complex and Nonlinear Phenomena (MCNP) research group members at Northumbria University and, in particular, Thibault Bonnemien (now at King's College), Thibault Congy and Gennady El for useful remarks. In particular, the authors are indebted with Gennady El for providing them with a physical copy of \cite{GurevichShvartsburg1973}. The authors would also like to thank Bob Jenkins, Sara Lombardo, Alexander Tovbis and Cornelis van der Mee for enlightening discussions. GO would like to thank Bulat Suleimanov for bringing \cite{Suleimanov2017} to his attention. GR and MS would like to thank the Isaac Newton Institute for Mathematical Sciences for support and hospitality during the Dispersive Hydrodynamics programme, when work on this paper was undertaken (EPSRC Grant Number EP/R014604/1). GR's work was also supported by the Simon Foundation and by the EPSRC Grant Number EP/W032759/1. GO's work was partly supported by the European Union's Horizon 2020 research and innovation programme under the Marie Sk{\l}odowska-Curie Grant Number 778010 IPaDEGAN. The work of GO was carried out within the framework of the MMNLP (Mathematical Methods in Non-Linear Physics) project of the INFN (Istituto Nazionale Fisica Nucleare). Finally, the work of FD, GO and MS has been carried out under the auspices of the Italian GNFM (Gruppo Nazionale Fisica Matematica), INdAM (Istituto Nazionale di Alta Matematica), that is gratefully acknowledged.


\end{document}